\documentclass[prd,preprint,superscriptaddress]{revtex4}
\usepackage{revsymb}
\usepackage{amsmath}
\usepackage{graphicx}
\usepackage{bm}
\newcommand{\be}{\begin{equation}}
\newcommand{\ee}{\end{equation}}
\newcommand{\ben}{\begin{eqnarray}}
\newcommand{\een}{\end{eqnarray}}

\begin{document}
\title{Interacting models may be key to solve the cosmic coincidence problem}
\author{Sergio del Campo\footnote{Electronic Mail-address:
sdelcamp@ucv.cl}} \affiliation{Instituto de F\'{\i}sica,
Pontificia Universidad Cat\'{o}lica de Valpara\'{\i}so, Avenida
Brasil 2950, Casilla 4059, Valpara\'{\i}so, Chile}
\author{Ram\'{o}n Herrera\footnote{E-mail address: ramon.herrera@ucv.cl}}
\affiliation{Instituto de F\'{\i}sica, Pontificia Universidad
Cat\'{o}lica de Valpara\'{\i}so, Avenida Brasil 2950, Casilla
4059, Valpara\'{\i}so, Chile}
\author{Diego Pav\'{o}n\footnote{E-mail address: diego.pavon@uab.es}}
\affiliation{Departamento de F\'{\i}sica, Facultad de Ciencias,
Universidad Aut\'{o}noma de Barcelona, 08193 Bellaterra
(Barcelona), Spain}

\begin{abstract}
It is argued that cosmological models that feature a flow of
energy from dark energy to dark matter may solve the coincidence
problem of late acceleration (i.e., ``why the energy densities of
both components are of the same order precisely today?"). However,
much refined and abundant observational data of the redshift
evolution of the Hubble factor are needed to ascertain whether
they can do the job.
\end{abstract}


\maketitle

\section{Introduction}
As is well known, the  popular vacuum-cold dark matter
($\Lambda$CDM) model fits rather well most observational data,
however it suffers from two main -and lasting- shortcomings,
namely: the low value of the vacuum energy (about 123 magnitude
orders below the quantum field theory estimation) and the
coincidence problem. The latter refers to the fact that despite
the vacuum energy does not vary with time and dark matter energy
depends on redshift as $(1+z)^{3}$ both energy densities are
nowadays of the same order. This is why very often the vacuum
component is replaced by an evolving dark energy  field -see
\cite{rreviews,padm} for recent reviews. Nevertheless, this does
not satisfactorily solve the problem since the initial value of
the dark energy ought to be fine tuned up to obtain the desired
result; it only substitutes a fine tuning by another \cite{padm}.

An encouraging approach that is receiving increasing attention
\cite{list} rests on the assumption that  both components do not
evolve separately but interact with each other, i.e.,
\begin{equation}
\dot{\rho}_{m} + 3 H \rho_{m}  = Q \, , \qquad {\rm and} \qquad
\dot{\rho}_{\phi} + 3 H (\rho_{\phi} + P_{\phi}) = -Q \, .
\label{cons}
\end{equation}
Here, $Q$ stands for the interaction term, and the subscripts $m$
and $\phi$ denote non-relativistic (pressureless) dark matter (DM)
and dark energy (DE), respectively. Clearly, this represents a
step forward because for $Q > 0$ (i.e., for energy transfers from
DE to DM) the ratio between both energies, $r \equiv
\rho_{m}/\rho_{\phi}$, evolves more slowly than in the
$\Lambda$CDM model. Effectively,  in the latter, $\dot{r} = -3H r$
while in the interacting case its time evolution of  governed by
\begin{equation}
\dot{r} = 3 H
r\left[w+\frac{\kappa^{2}\,Q}{9\,H^3}\frac{(r+1)^2}{r} \right] \,
, \label{rr}
\end{equation}
where $ w \equiv P_{\phi}/\rho_{\phi} < -1/3$ denotes the equation
of state parameter of DE. Equation (\ref{rr}) follows from
(\ref{cons}) and Friedmann's equation for spatially flat universes
\begin{equation}
3 H^2 = \kappa^{2} (\rho_{m} + \rho_{\phi}) \quad
\qquad(\kappa^{2} \equiv 8\pi\, G)\, . \label{Fried1}
\end{equation}
As we will see below, for reasonable and suitably chosen
interaction terms, $r$ either tends to a constant or varies more
slowly than the scale factor, $a(t)$, at late times. This
certainly alleviates the coincidence problem. It is worthy of note
that if $Q$ were negative (signifying an energy transfer from DM
to DE) the problem would only worsen, moreover at early times the
energy density of DE would have been negative, and, finally, the
second law of thermodynamics could be violated \cite{lima,db}.
Except in chamaleon models \cite{chamaleon}, baryons as well as
radiation usually do not participate in the interaction. The
former because of the tight constraints imposed by local gravity
measurements \cite{Peebles2,Hagiwara}, the latter because
otherwise the photons would not longer follow geodesics which
would affect precise measurements of deviations of radar signals
grazing the sun.

It is not the aim of this article to solve the coincidence problem
-we believe it cannot be done at present- but to devise a simple
strategy to do it, within general relativity, as soon as
sufficiently accurate and abundant data on the evolution of the
Hubble factor becomes available. This paper extends and
generalizes a previous work of the authors \cite{rapid}.

Most studies of late acceleration implicitly assume that matter
and dark energy interact gravitationally only. In the absence of
an underlying symmetry that would suppress a matter-dark energy
coupling (or interaction) there is no a priori reason to discard
it. Further, the coupling is not only likely but inevitable
\cite{jerome} and its introduction is less arbitrary than
otherwise. In fact, the accretion of DE by black holes
\cite{cai-wang,bean} implies a non-vanishing and positive $Q$.
While its small value modifies only very little the dependence of
the energy densities on redshift (and cannot substantially
alleviate the coincidence problem), it sets a minimum value for
the interaction.

The interaction may bring some further consequences. It can push
the beginning of accelerated expansion era to higher redshifts. It
can also  erroneously suggest an equation of state for the dark
energy of phantom type (i.e., $w <-1$), see
\cite{amendola-gasperini} and references therein. Moreover,  the
interaction may give rise to fluctuations in the count of galaxy
clusters with redshift \cite{mota}. Likewise, the coupling can
alter the isothermal Maxwell-Boltzmann velocity distribution of
weakly interacting massive particles in the galaxy halos
\cite{Tetradis1} whereby the average dark matter velocity is
expected to increase significantly. Since the detection rates of
the different experiments searching for dark matter strongly
depend on this velocity it is crucial not to dismiss {\em a
priori} the coupling \cite{Tetradis2}.

Ultimately, the existence or non-existence of the interaction is
to be discerned observationally. Whereas the current empirical
data are insufficient to settle this issue, a couple of studies
seem to favor the former possibility: $(i)$ As it should  be
expected, the interaction alters the time required for a
self-gravitating, collapsing, structure to reach equilibrium as
well as the equilibrium configuration itself. Therefore the
Layzer-Irvine equation \cite{layzer,Peebles1} is to be generalized
to take into account the interaction. In this regard, from the
analysis of the dynamics of 33 relaxed galaxy clusters -for which
reliable x-ray, weak lensing and optical data are available-, it
has been reported a small but not vanishing interaction
\cite{elcio,orfeu}. $(ii)$ Since the interaction modifies the
evolution rate of the metric potentials the integrated Sachs-Wolfe
component of the cosmic microwave background radiation (CMB) is
enhanced. The cross-correlation of galaxy catalogs with CMB maps
also hints at a small interaction \cite{isw-olivares}.

This paper is organized as follows. Next section introduces three
different expressions for $Q$ and show that all three lead to a
late, everlasting, era of either constant or quasi-constant $r$
regardless of whether $w$ is lower or higher than $-1$. Section 3
shows the consistency of the three interacting models  with the
recent observational bounds of Daly {\it et al.} \cite{daly1}
based on data from supernovae type Ia, taken from different
sources, and 30 radio-galaxies. The only cosmological assumption
behind these bounds is that the  Universe can be considered
homogeneous and isotropic at large scales. Section 4 devises a
simple strategy to solve the coincidence problem. Finally, Section
5 summarizes and discusses our results.

We focus our attention on spatially flat
Friedmann-Robertson-Walker models. As usual, a subscript zero
means that the corresponding quantity is to be evaluated at
present time.

\section{Three choices for the interaction term}
Because of the unknown nature of both DM and DE it does not seem
feasible nowadays to derive an expression for $Q$ on first
principles. The alternative is to propose phenomenological
expressions based on arguments of simplicity and reasonableness.
Firstly, note that $Q$ must be small and positive -otherwise the
DE could not dominate the expansion today. On the other hand, if
it were large and negative, DE would have dominated the expansion
practically from the outset and galaxies couldn't have condensed.

Looking at the left hand side of Eqs. (\ref{cons}) it becomes
apparent that $Q$ must be a function of the energy densities
multiplied by a quantity with units of inverse of time. For the
latter the obvious choice is the Hubble factor, so  $Q=
Q(H\rho_{m}, \, H \rho_{\phi})$. By power law expanding this
function and keeping just the first term we get $Q \simeq
\lambda_{m} \, H \,  \rho_{m} + \lambda_{x} \, H \, \rho_{x}$. In
view of the lack of information about the coupling it is advisable
to retain just one parameter (the less observational parameters
when fitting to observational data, the better). Then, three
choices follow, namely: $\lambda_{m} = \lambda_{x}$, $\lambda_{x}
= 0$, and $\lambda_{m} = 0$. Other choices, as $Q \propto \rho_{m}
\,\rho_{x}$ \cite{binw}, are also possible but we will not deal
with them here.

We will therefore consider the following three expressions for the
coupling,
\begin{equation}
Q = 3 \alpha H (\rho_{m} +\rho_{\phi}) \, ,  \; \qquad Q = 3 \beta
H \, \rho_{m} \, , \; \qquad {\rm and} \,  \; \qquad Q = 3 \eta H
\rho_{\phi} \, , \label{3Q}
\end{equation}
where the phenomenological parameters, $\alpha$, $\beta$, and
$\eta$, are dimensionless, positive-definite small constants.
Henceforward, we will call the corresponding models alpha, beta,
and eta, respectively.

Model alpha fits very well data from SN Ia, CMB, and large scale
structure formation provided $\alpha < 2.3 \times 10^{-3}$
\cite{olivares1,olivares2,isw-olivares}. Further, this model
possesses the remarkable property that the ratio $r$ tends to a
stationary but unstable value at early times, $r^{+}_{s}$, and to
a stationary and stable value, $r^{-}_{s}$, (an attractor) at late
times. This follows from inserting the corresponding expression
for $Q$ in the right hand side of Eq. (\ref{rr}) and setting
$\dot{r}$ to zero. For $w =$ constant the stationary solutions of
the resulting quadratic equation are:
\begin{equation}
r^{\pm}_{s} = -1 + 2b \pm 2 \sqrt{b(b-1)}\, ,  \qquad b = -
\frac{w}{4\, \alpha} > 1.
\label{r+-}
\end{equation}

\begin{figure}[th]
\includegraphics[width=7.0in,angle=0,clip=true]{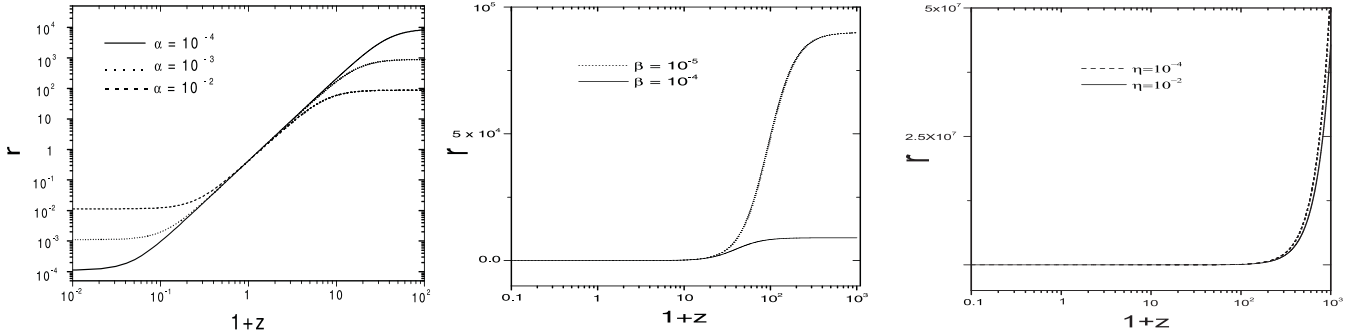}
\caption{From left to right, evolution of the ratio
$r=\rho_{m}/\rho_{\phi}$ with redshift for  models alpha, beta,
and eta. For all of them $r$ either tends to a constant or varies
very slowly at small redshift. In drawing the curves we have fixed
$r_{0}= 3/7 \,$ and $\, w= -0.9$. } \label{fig:quinteshor}
\end{figure}

\begin{figure}[th]
\includegraphics[width=7.0in,angle=0,clip=true]{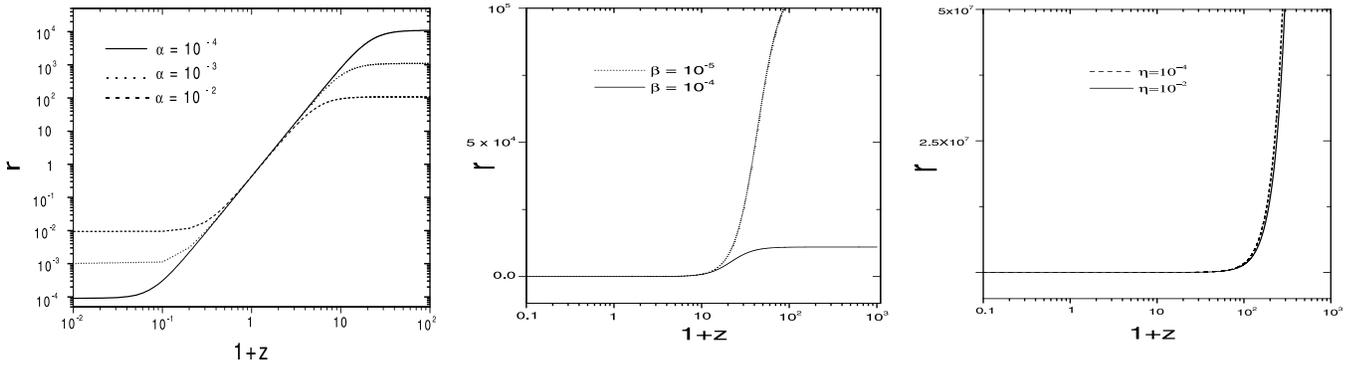}
\caption{Same as Fig.1 except for that in this case $w= -1.1$. }
\label{fig:phantomhor}
\end{figure}

As it can be checked by slightly perturbing the right hand side of
Eq.(\ref{rr}), the stationary solution  $r^{+}_{s}$ results
unstable while $r^{-}_{s}$ is stable \cite{plbwdl,prdladw}. The
general solution of Eq. (\ref{rr})
\begin{equation}
r(x) = \frac{r^{-}_{s}+ x r^{+}_{s}}{1 + x}\, ,
\label{r(x)}
\end{equation}
interpolates between $r^{+}_{s}\, $ and $\, r^{-}_{s}$. Here, $x =
(a/a_{*})^{- \mu}$, with $\mu \equiv 12\, \alpha \sqrt{b(b-1)}$,
and $a_{*}$ denotes the scale factor at which $r$ takes the
arithmetic medium value $(r^{+}_{s}+ r^{-}_{s})/2$. In the range
$r^{-}_{s} < r < r^{+}_{s}$ the function $r(x)$ decreases
monotonously. Consequently, as the Universe expands, $r(x)$
smoothly evolves from $r_{s}^{+}$ to the attractor solution
$r^{-}_{s}$. The transition from one asymptotic solution to the
other occurred only recently -see left panels of Figs.
\ref{fig:quinteshor} and \ref{fig:phantomhor}- so one can take
$r\simeq r^{+}_{s}$ during a fairly large part of the history of
the Universe. Finally, the constraint $r_{s}^{+} \simeq$ constant
implies that $r$ and $\alpha$ are not independent, but verify the
relationship $\alpha (r_{s}^{+}\, +1)^2= \mid w \mid \,
r_{s}^{+}$, so the product $\alpha \, r_{s}^{+} \sim \mid w \mid$
is of order unity.

As indicated above, this choice of $Q$ yields a constant but
unstable ratio at early times and constant and stable lower ratio
at late times -left panels of Figs. \ref{fig:quinteshor} and
\ref{fig:phantomhor}. It is hard to imagine a simpler expression
for the coupling entailing these two properties. Likewise, the
aforesaid expression can be re-interpreted as implying an
effective exponential potential for the dark energy field at late
times and a power-law at early times \cite{plbwdl,olivares1}.

Near  $r \approx r^{-}_{s}$ the balance Eqs. (\ref{cons}) can be
approximated by
\begin{eqnarray}
\frac{1}{\rho_{m}}\, \frac{d \rho_{m}}{d r} &\simeq &
\frac{1-\alpha (1+1/r^{-}_{s})}{\alpha (r^{+}_{s} - r^{-}_{s}) (r -r^{-}_{s} )} \, ,
\nonumber \\
\frac{1}{\rho_{\phi}}\, \frac{d \rho_{\phi}}{d r} & \simeq &
\frac{1+w + \alpha (1+r^{-}_{s}}{\alpha (r^{+}_{s} - r^{-}_{s}) (r
-r^{-}_{s})} \, \, . \label{near-attractor}
\end{eqnarray}
For $w \simeq$ constant, these equations can be integrated to
\begin{equation}
\rho_{m} \propto a^{-3\left[ 1- \alpha(1+ 1/r^{-}_{s})\right]}\, ,
\qquad \rho_{\phi} \propto a^{-3\left[1+w + \alpha (1+r^{-}_{s})%
\right]} \, .
\label{integrate}
\end{equation}
Notice that the condition $\dot{r}|_{r = r^{-}_{s}} = 0$ implies
that the exponents in the energy densities, Eq. (\ref{integrate}),
coincide.  Interestingly, these results are not only valid when
the dark energy is a quintessence field (i.e., $-1 < w <-1/3 $),
they also apply when the dark energy is of phantom type (i.e., $w
<-1$), either a scalar field with the ``wrong sign" for the
kinetic energy term, a {\em k}-essence field, or a tachyon field
\cite{dualk}.

In the vicinity of the attractor, dark matter and dark energy
dominate the expansion whence the scale factor can be written as
\[
a \propto t^{(2/3)[1+w + \alpha (1+r^{-}_{s})]^{-1}}.
\]

It is easy to show that, at early times, $r^2+(w\,
\alpha^{-1}+2)r+1=0$ and the constant ratio is $r_{s}^{+}
=r(z\rightarrow\infty)\sim \alpha^{-1}$ while, at late times,
$r_{s}^{-}=r(z\rightarrow -1)\sim \alpha $ \cite{prdladw}. These
results mean a conceptual advance with respect to the $\Lambda$CDM
model. If the present acceleration of the Universe is generated by
a cosmological constant, then the initial values of dark matter
and dark energy have to be fine-tuned by 96 orders of magnitude at
Planck time, i.e., $r(t_{Planck})=\Omega_{m}/\Omega_{\Lambda}\sim
{\cal O}(10^{96})$. In the model of above this ratio is fixed by
$\alpha$. Thus, the initial condition problem is significantly
alleviated. For example, if $\alpha \sim 10^{-4}$, the initial
dark matter to dark energy ratio comes to be
$r(t_{Planck})\sim{\cal O}(10^4)$.

Model beta does not share the nice feature of a late attractor.
However, in this case,  for constant $w$ one has \cite{softk}
\begin{equation}
r= \frac{\xi}{\beta r_{0}-(1+z)^{-3\xi}\, [\beta (1+r_{0})+w]}\,
r_{0} \, , \label{r}
\end{equation}
where $\, \xi = -w-\beta > 0$. Hence,  the evolution of $r$ is
slower than in the $\Lambda$CDM model. There is a monotonous
decrease of $r$ from values higher than unity at early times to a
nearly constant, value at present time (fixed as $3/7$)
irrespective of whether dark energy is a quintessence field
(middle panel of Fig. \ref{fig:quinteshor})  or a phantom field
(middle panel of Fig. \ref{fig:phantomhor}). For vanishing scale
factor $r$ tends to the finite, constant value $\xi/\beta$
(however, the model should not be extrapolated to such an early
stage). Thus, in spite of the absence of a late time attractor
this choice of $Q$ also significantly alleviates the coincidence
problem as the energy density ratio varies very slowly at present
time, i.e., $\mid(\dot{r}/r)_{0} \mid \leq H_{0}$. This feature is
shared by model eta -see right panel in Figs. \ref{fig:quinteshor}
and \ref{fig:phantomhor}.

Altogether, all three models considerably alleviate the
coincidence problem but there is no guarantee that any of  them
solves it in full. A successful model should also predict $r_{0}
\sim {\cal O}(1)$. To the best of our knowledge, no model in the
market does this. We believe that, for the time being, the present
value of $r$ must be considered an input parameter much in the
same way as the current value of the Hubble function, $H_{0}$, the
temperature of the CMB, the age of the universe,  or the ratio
between the number of baryons and photons, are -only that the four
latter are not laden with any puzzle.

\section{Comparison with observational data}
We wish to explore whether the interacting, models alpha, beta,
and eta, are consistent with the results of Daly {\it et al.}
\cite{daly1}. These are based in the determination of the
dimensionless coordinate distance, $y(z) = H_{0}\, (a_{0}
\tilde{r})$,  and their two first derivatives with respect to
redshift, of supernovae type Ia (SN Ia) and radio-galaxies (RG) in
the redshift interval $0 < z < 1.8$. The supernovae samples
contains 182 SN Ia of Riess {\it et al.} \cite{riess}, 192 of
Davis {\it et al.} \cite{davis} and 115 of Astier {\it et al.}
\cite{astier} though there is some overlap between the different
samples. The RG sample comprehend 19 of Daly {\it et al.}
\cite{daly2} plus 11 additional RG.

It is to be stressed that the determination of the dimensionless
coordinate distances does not involve any cosmological model, just
the Friedmann-Robertson-Walker (FRW) metric. The first and second
derivatives of the dimensionless coordinate distance with respect
to redshift can be used to build a model-independent determination
of the dimensionless expansion rate, $E(z)=H(z)/H_0$, and the
deceleration parameter, $q(z)=-\ddot{a}/(a \, H^{2})$, see
\cite{djorgovski}.

Indeed, in virtue of the relation between redshift $z$ and the
scale factor $1+z= a_{0}/a(t)$, the definition of the Hubble rate,
$H = \dot{a}/a$, and the relationship $dt = -a(t) \, d\tilde{r}$,
valid in spatially flat FRW spacetimes for photons flying from
their source to the observer, one follows:
\begin{equation}
E(z)=\frac{H(z)}{H_0}=\frac{1}{y'(z)}\, ,
\label{E}
\end{equation}
where  a prime indicates derivation with respect to $z$. From last
expression the data $y(z)$ can be used to empirically determine
$y'(z)$ and $E(z)$. Differentiating  Eq.(\ref{E}) with respect to
time  and using the definition of $q(z)$ leads to
\begin{equation}
q(z)=- 1-(1+z)\, \frac{y''}{y'} \, . \label{q}
\end{equation}

Likewise, differentiating  Eq.(\ref{Fried1}) with respect to time
and using Eqs.(\ref{E}) and (\ref{q}) produces
\begin{equation}
P_\phi(z)=P_\phi(z=0)\,\left(\frac{E^2(z)}{E^2(z=0)}\right)\,\left[\frac{2\,q(z)-1}{2\,q(z=0)-1}\right].
\end{equation}

Thus, the dimensionless quantities $E(z)$,  $q(z)$ and
$P_{\phi}(z)$ can be constructed directly from $y'$ and $y''\,$
\cite{djorgovski}.

In virtue of  Eqs.(\ref{cons}),(\ref{Fried1}), (\ref{E}) and
(\ref{q}) we get
\begin{equation}
y'(z)=\frac{1}{\left[\Omega_{m0}\,\Im_1+\Omega_{\phi
0}\,\Im_2\right]^{1/2}}\, , \label{dy}
\end{equation}
and
\begin{equation}
y''(z)=-\frac{3}{2}\,\frac{y'}{(1+z)}\,\left[\frac{\Omega_{m0}\,\Im_1+\Omega_{\phi
0}\,(1+w)\, \Im_2} {\Omega_{m0}\, \Im_1+\Omega_{\phi 0}\,
\Im_2}\right]\, .
\end{equation}
These expressions are valid for all three models of above. Here
$\Omega_{m0}= \frac{\kappa^{2} \, \rho_{m0}}{3H^{2}}$ and
$\Omega_{\phi 0}= \frac{\kappa^{2} \, \rho_{\phi 0}}{3H^{2}}$
denote the present day values of the density parameters of DM and
DE, respectively. Numerical integration of Eq. (\ref{dy}) yields
the dimensionless coordinate distance, $y(z)=\int_0^z\,y'(z)\,dz$.
In doing this, $w =$ constant is assumed.

Next, we consider the three  expressions for the source term $Q$
introduced above.

$(i)$ For model alpha (interaction term
$Q=3\,\alpha\,H\,(\rho_{\phi}+\rho_{m})$), we find that
\begin{equation}
\Im_1(z)=(1+z)^{n}\, , \quad {\rm and}\, \quad
\Im_2=(1+z)^{n+m}\,,
\label{imalfa}
\end{equation}
with $n=\frac{3}{2}\,[2+w-\sqrt{w(w+4\alpha)}]$,  and
$m=3\sqrt{w(w+4\alpha)}$.

$(ii)$ For model beta ($Q=3\beta\,H\,\rho_{m}$) we obtain
\begin{equation}
\Im_1(z)=(1+z)^{n}\, , \quad {\rm and}\, \quad
\Im_2=(1+z)^{3(1+w)}\,-\frac{\beta}{(\beta+w)}\,
\frac{\Omega_m}{\Omega_{\phi}} \, \Im_1(z),
\label{imbeta}
\end{equation}
with $n=3(1-\beta)$.

$(iii)$ Likewise, for model eta ($Q=3\eta\,H\,\rho_\phi$)
\begin{equation}
\Im_1(z)=(1+z)^{3}-\frac{\Omega_\phi}{\Omega_m}\,\frac{\eta}{(\eta+w)}\,\Im_2(z)\,\,\;\;\;{\rm
and}\,\;\; \Im_2=(1+z)^{3(1+w+\eta)}.
\label{imeta}
\end{equation}

In the absence of interaction expressions (\ref{imalfa}),
(\ref{imbeta}), and (\ref{imeta}) reduce to Eqs. (1) and (2) of
Daly {\it et al.} \cite{daly1}.

\begin{figure}[th]
\includegraphics[width=6.in,angle=0,clip=true]{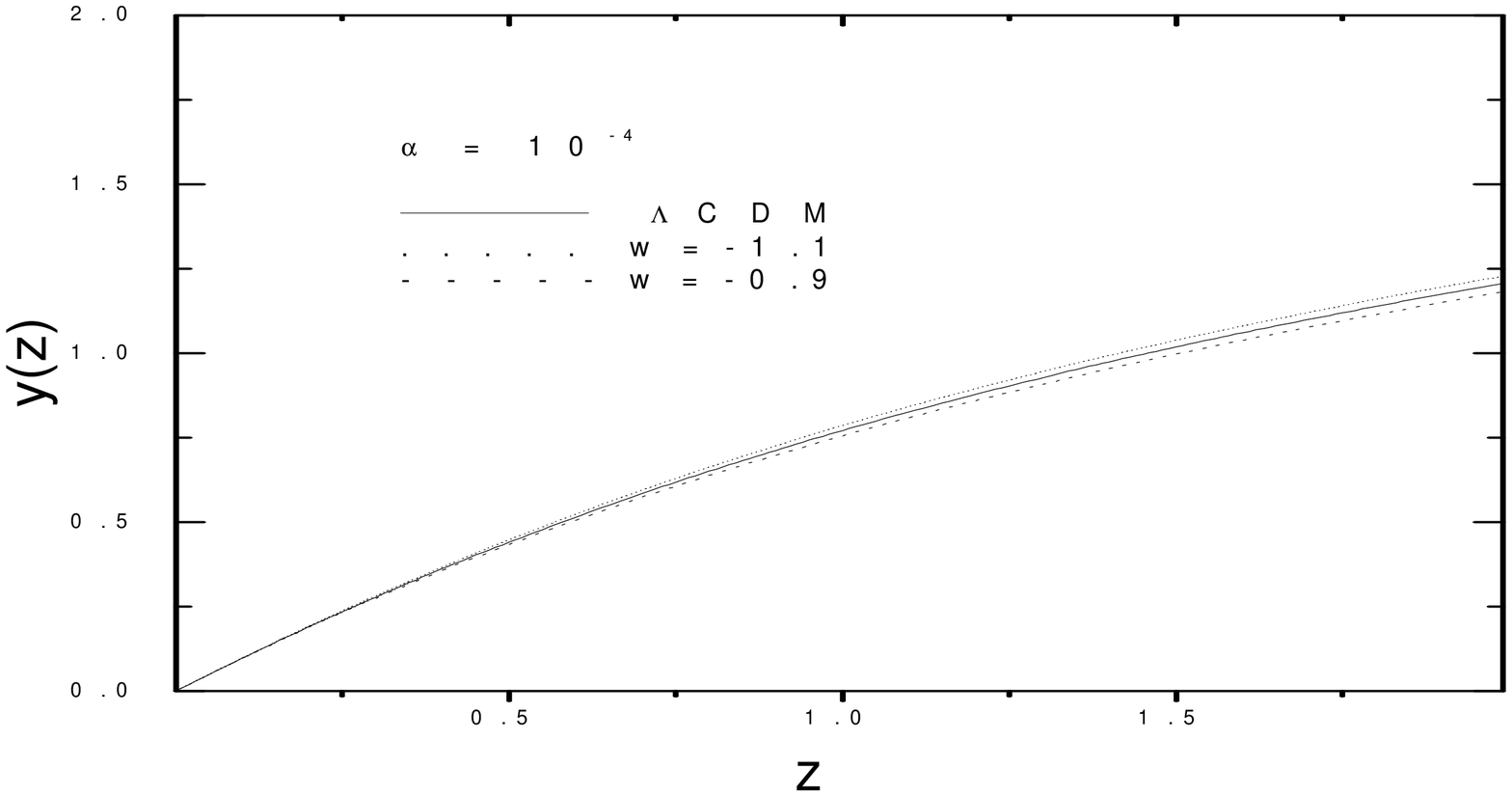}\vspace{-5.cm}
{\vspace{-3.cm}\includegraphics[width=6.
in,angle=0,clip=true]{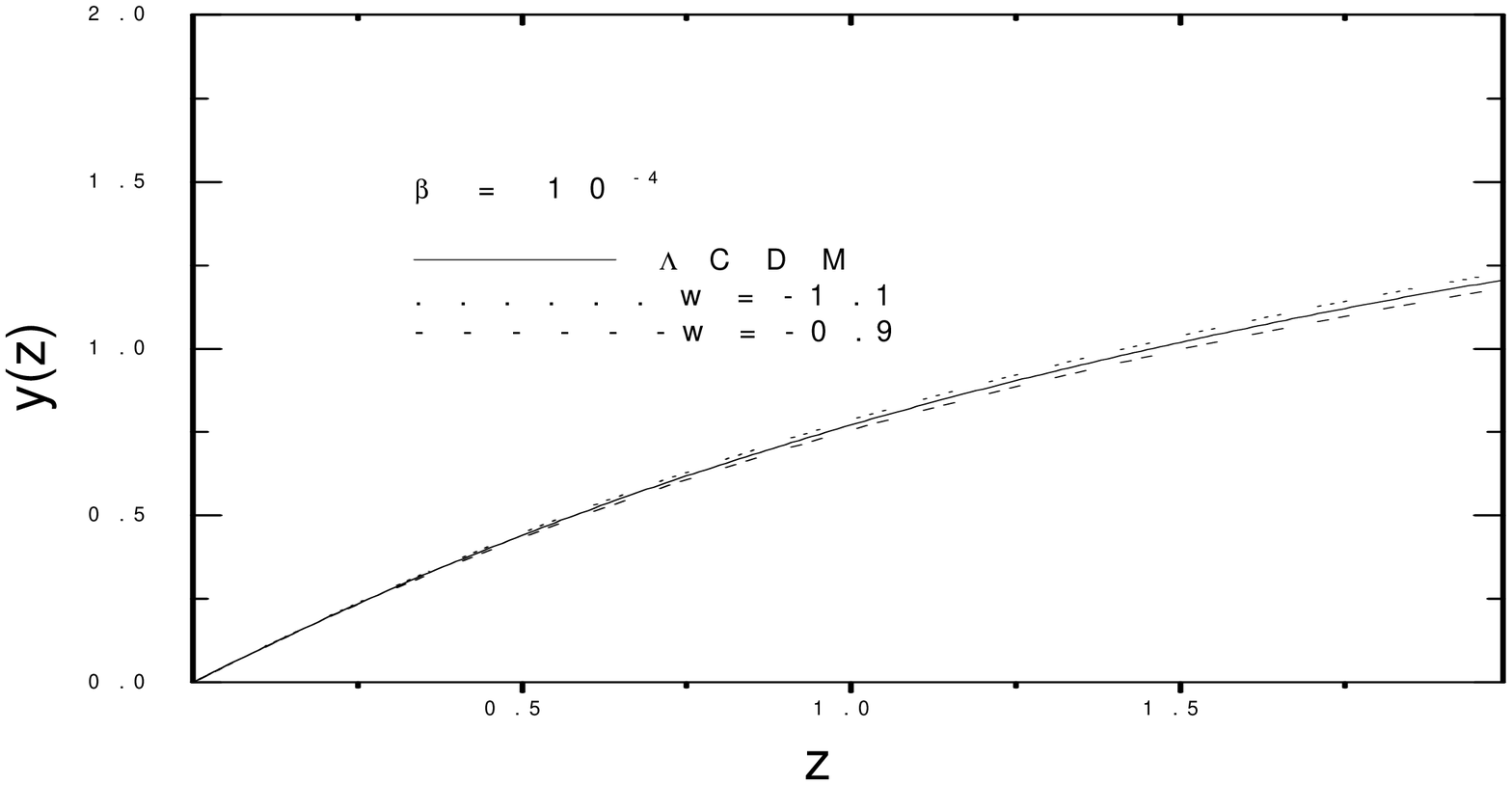}}\vspace{-2.cm}
{\vspace{-3.cm}\includegraphics[width=6.
in,angle=0,clip=true]{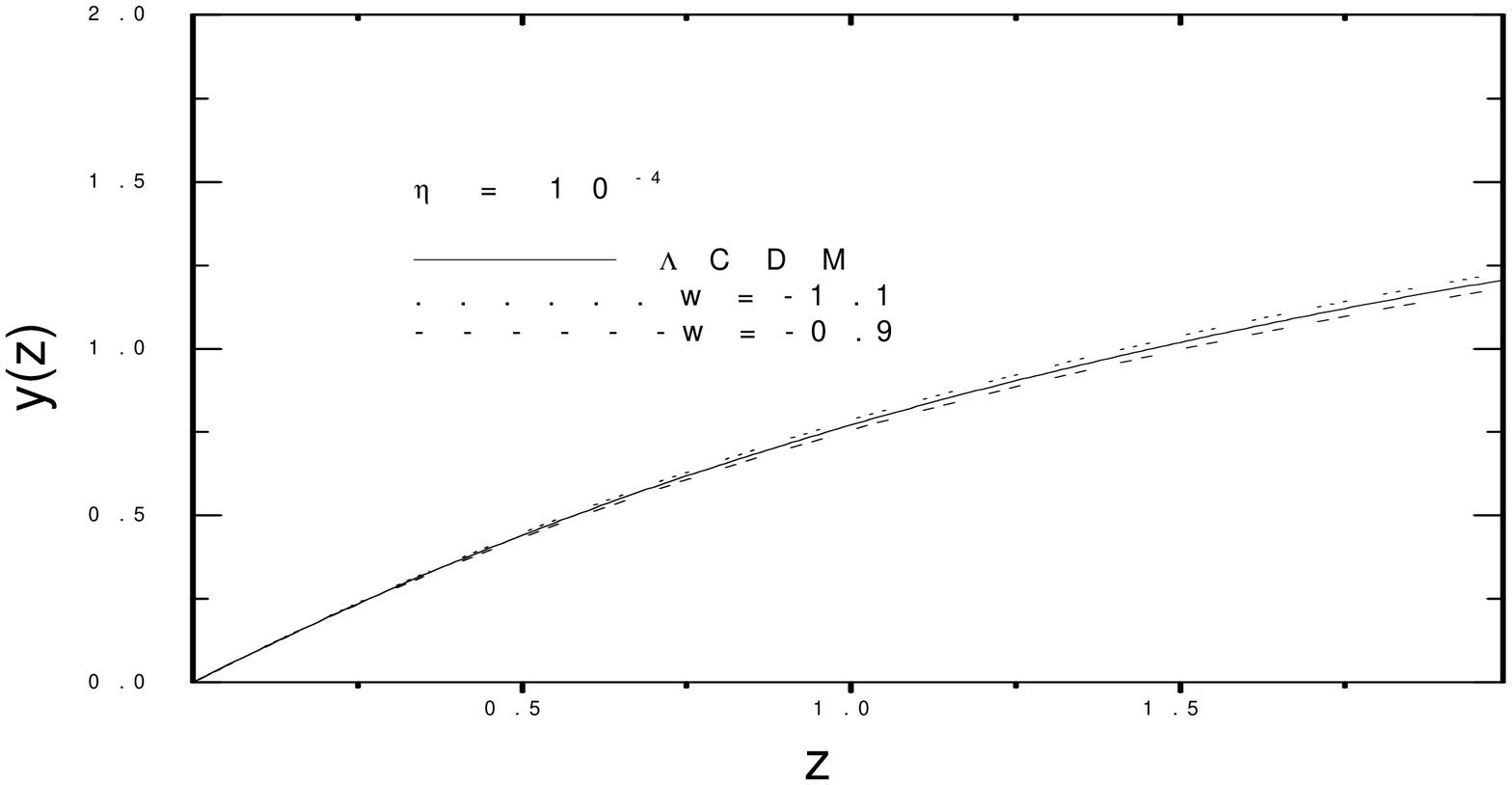}} \caption{From top to bottom,
dimensionless coordinate distance, $y(z) = H_{0} \, (a_{0}\,
\tilde{r})$, in terms of the redshift for models alpha, beta, and
eta. For comparison, the prediction of the $\Lambda$CDM model is
also shown.} \label{fig:rvsz2y}
\end{figure}

\begin{figure}[th]
\includegraphics[width=6.in,angle=0,clip=true]{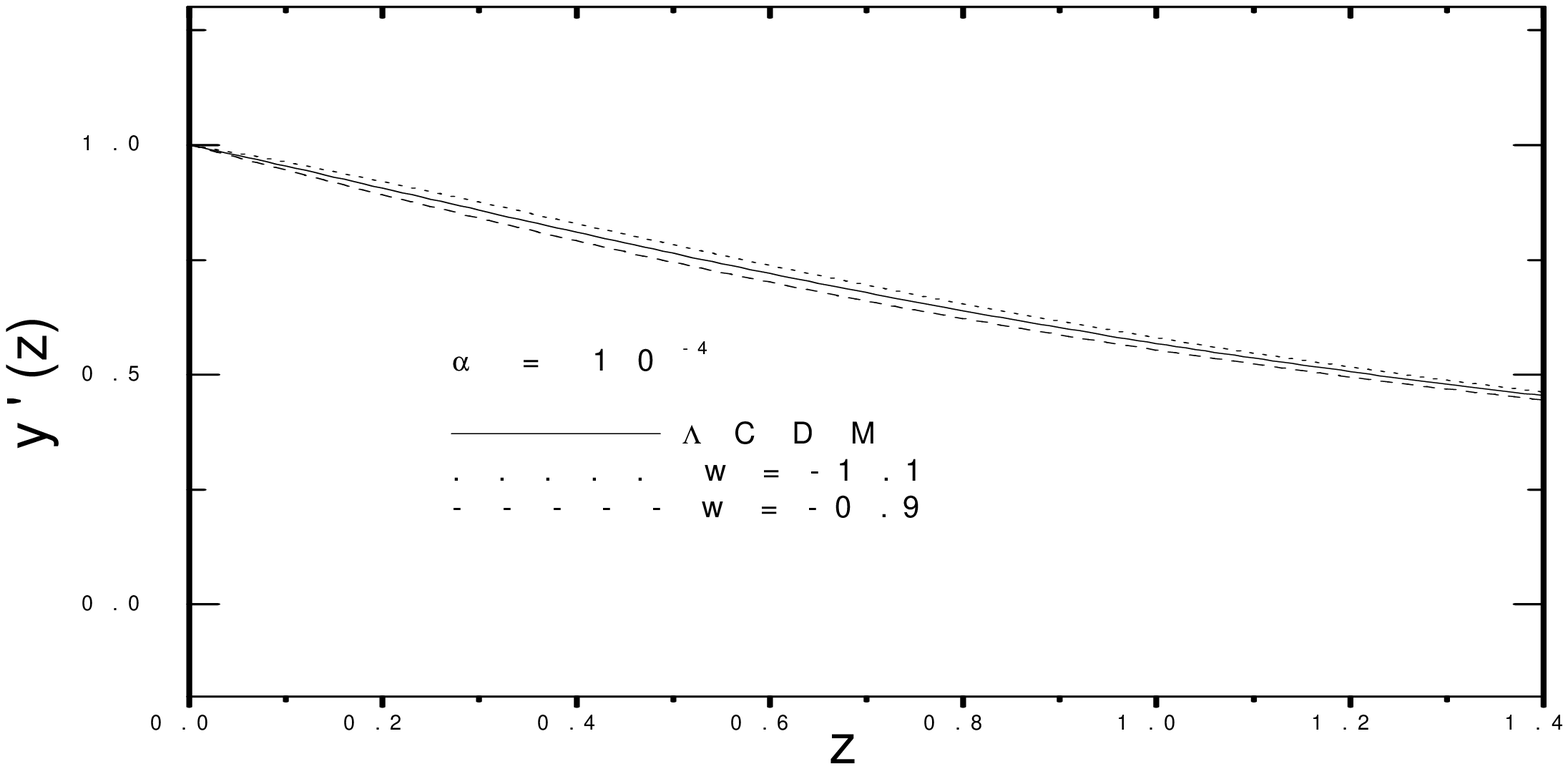}\vspace{-4.cm}
\vspace{-3.cm}\includegraphics[width=5.5in,angle=0,clip=true]{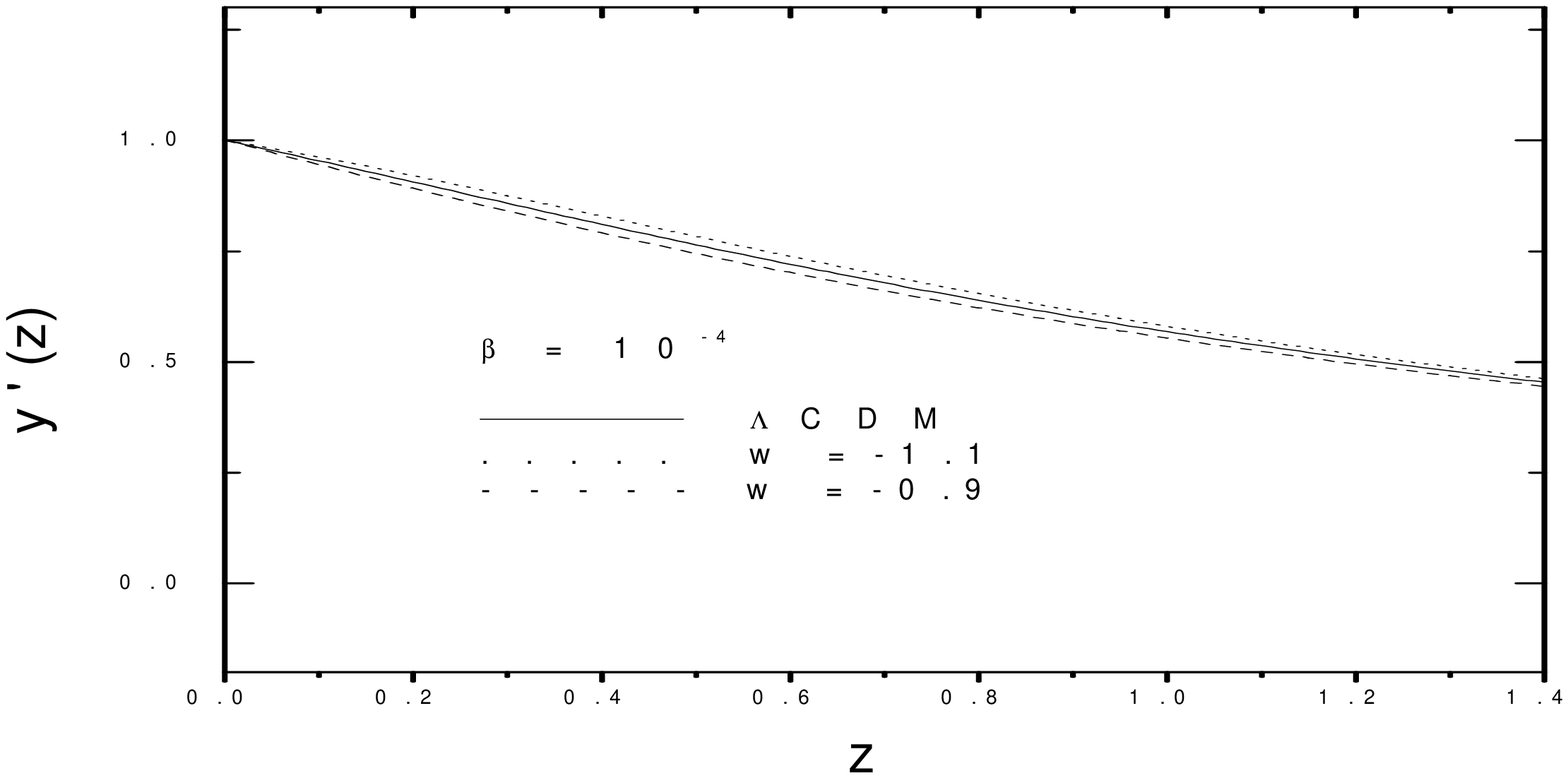}\vspace{-1.cm}
\vspace{-3.cm}\includegraphics[width=5.5in,angle=0,clip=true]{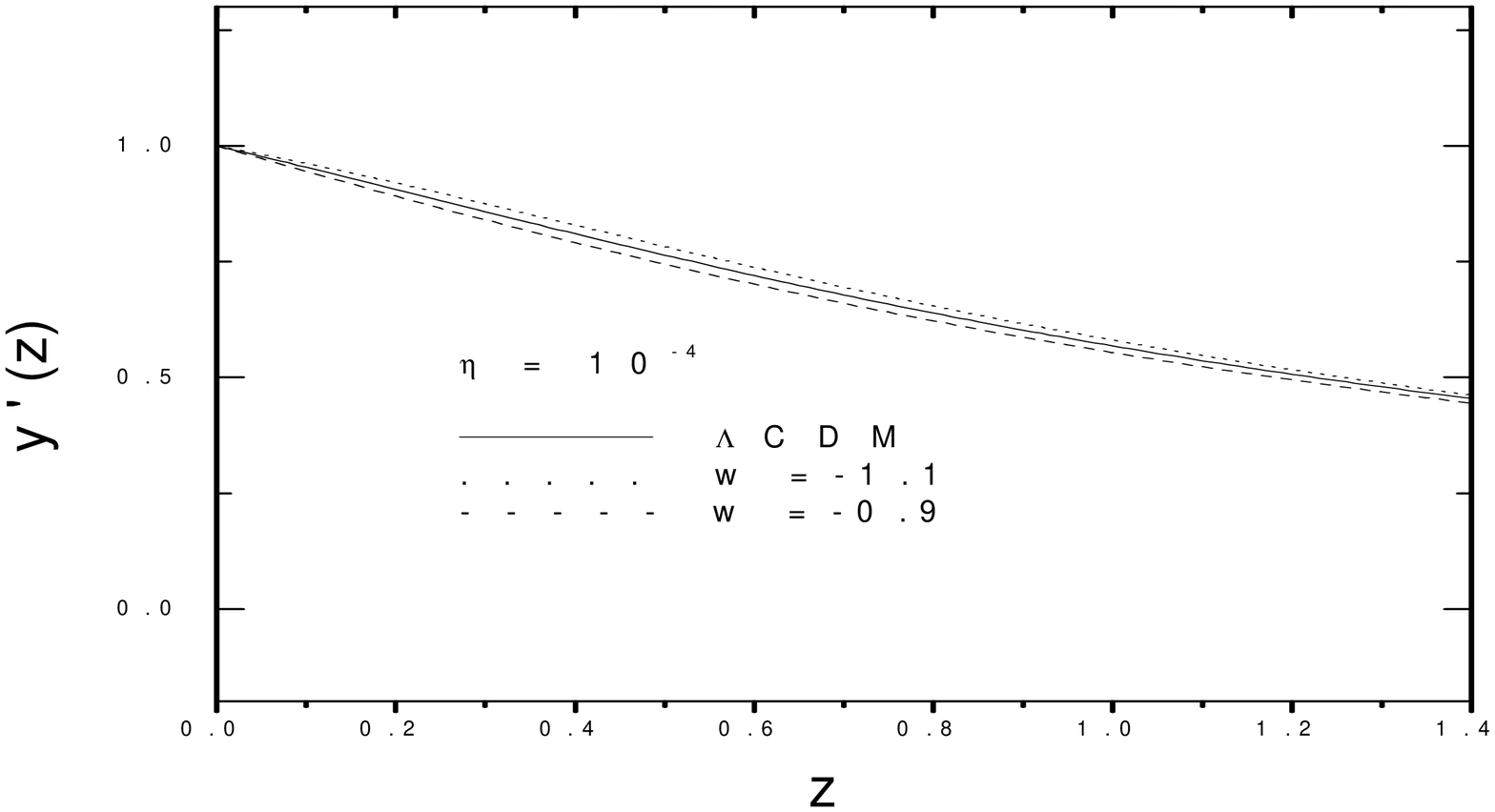}
\caption{From top to bottom, first derivative of the dimensionless
coordinate distance in terms of redshift for models alpha, beta,
and eta. For comparison, the prediction of the $\Lambda$CDM model
is also shown.} \label{fig:rvsz2dy}
\end{figure}

\begin{figure}[th]
\includegraphics[width=6.in,angle=0,clip=true]{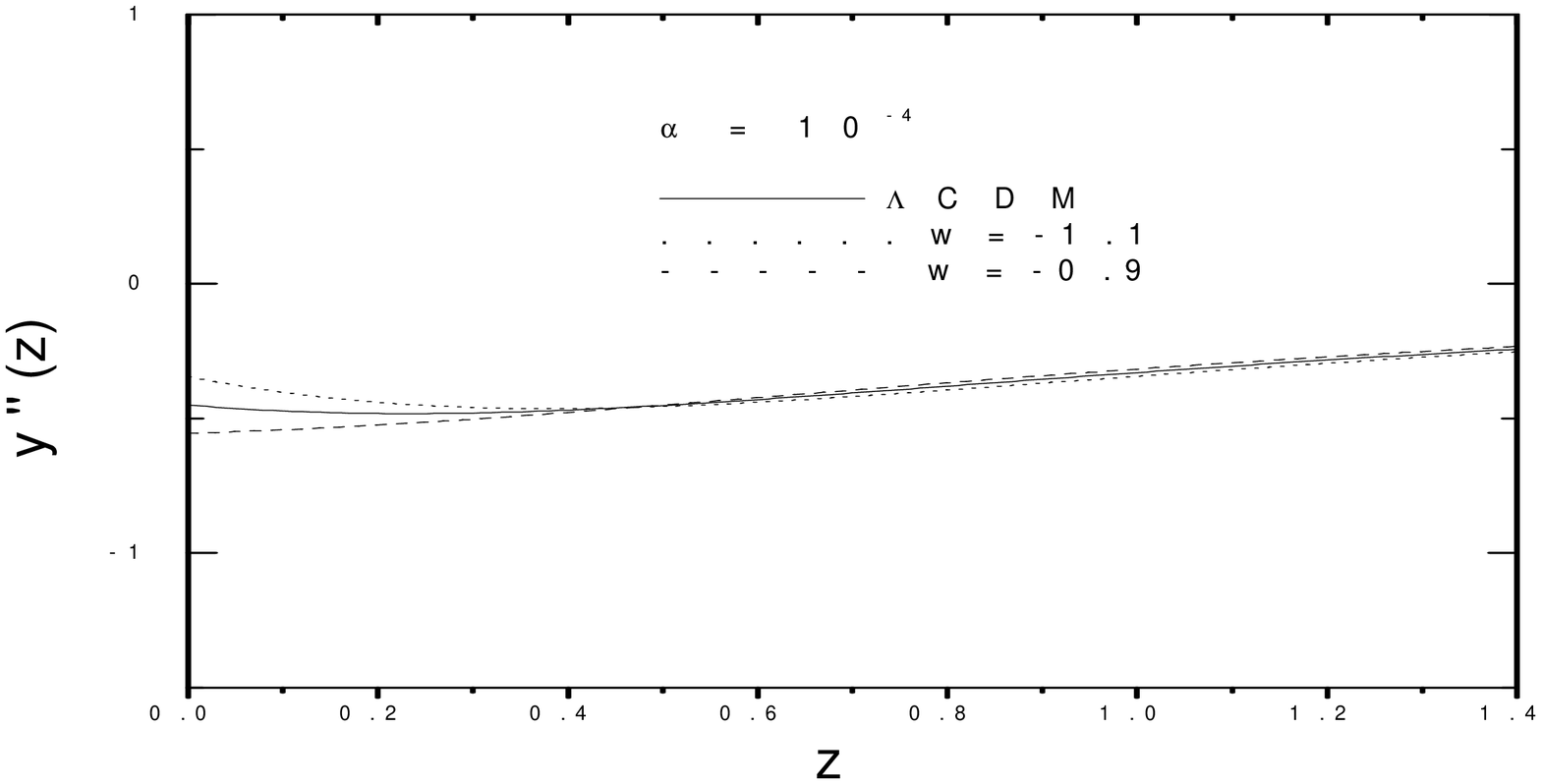}\vspace{-5.cm}
{\vspace{-3.cm}\includegraphics[width=6.in,angle=0,clip=true]{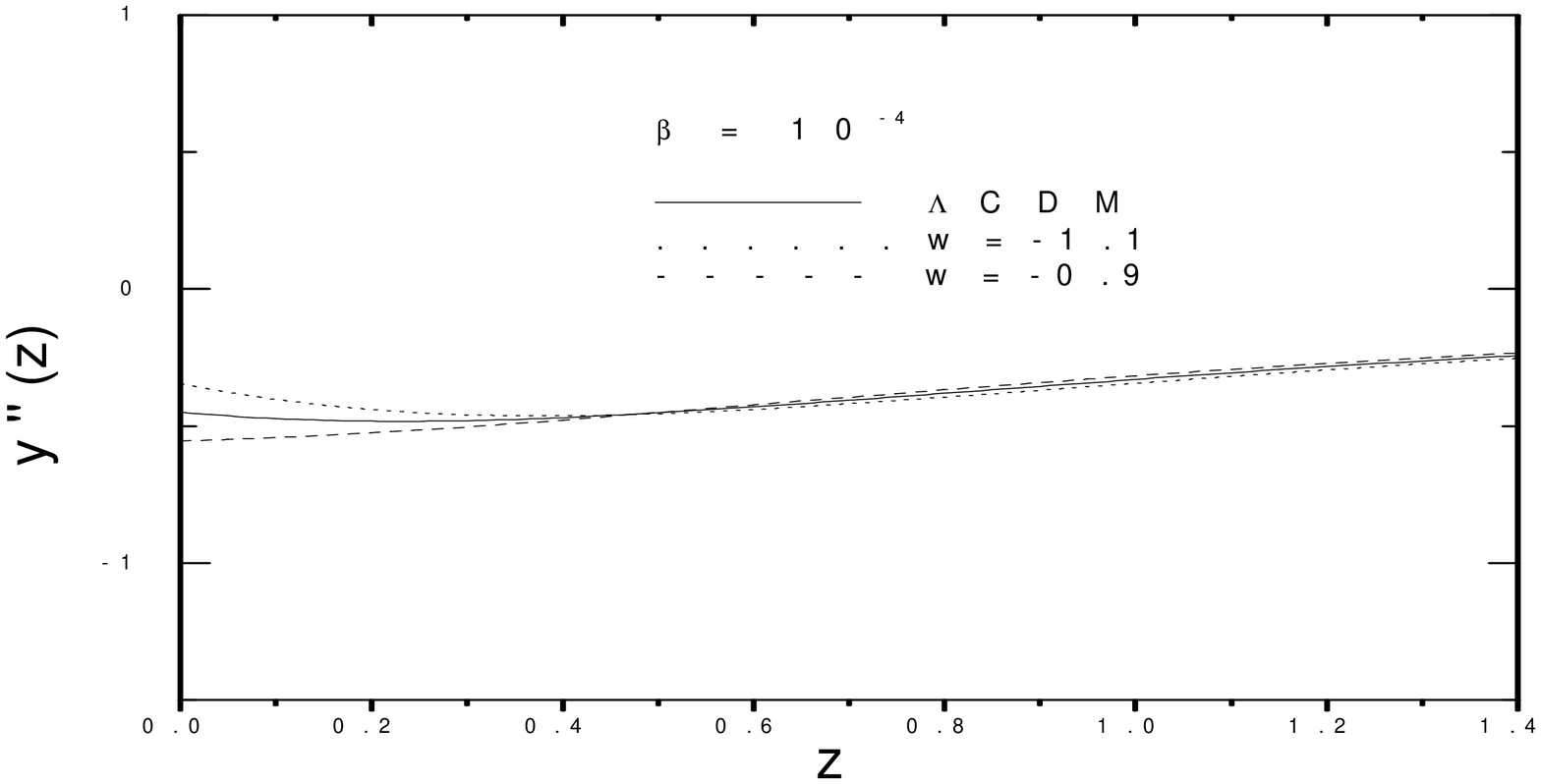}}\vspace{-1.6cm}
{\vspace{-4.cm}\includegraphics[width=6.in,angle=0,clip=true]{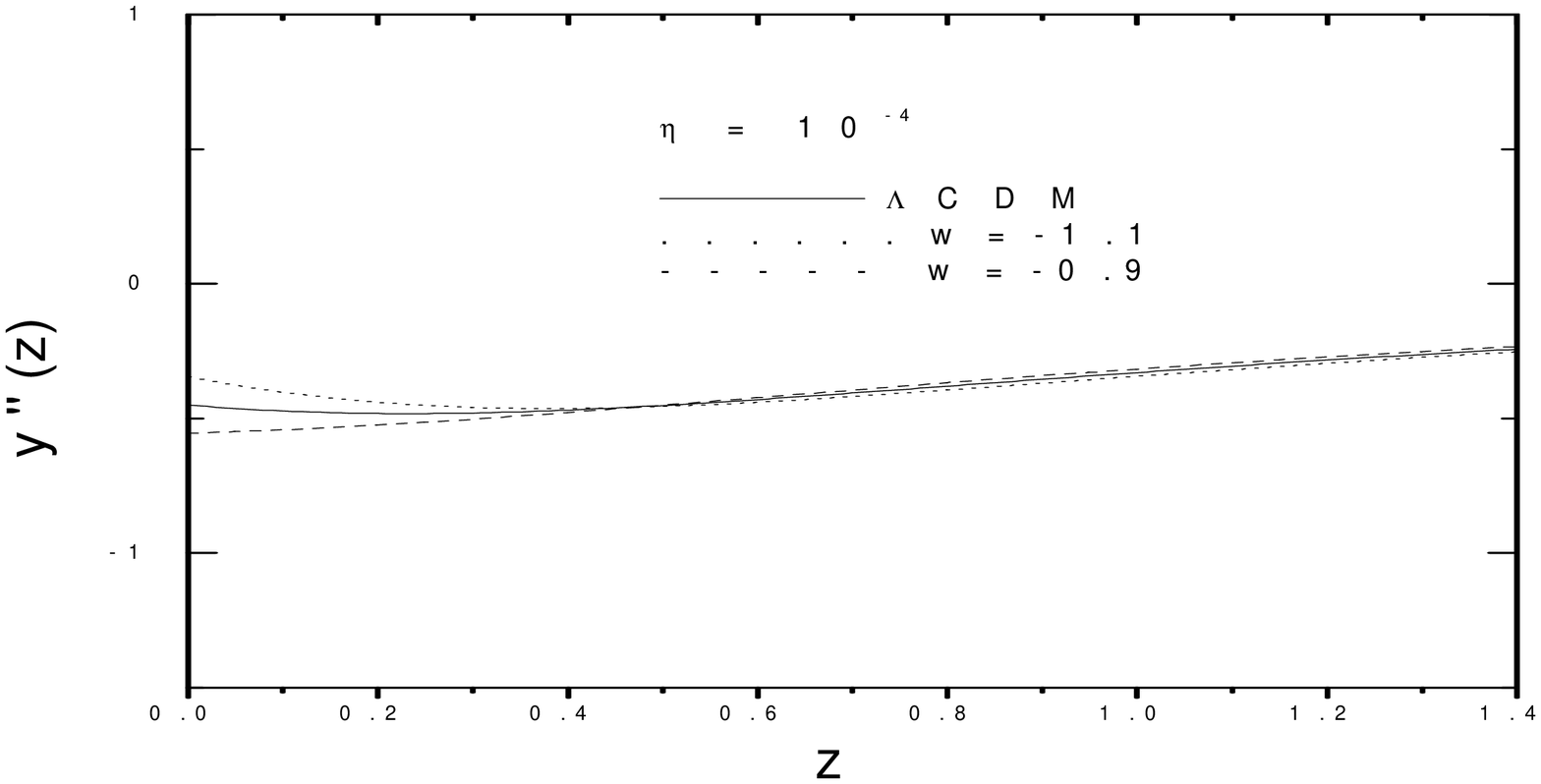}}
\caption{From top to bottom, second derivative of the
dimensionless coordinate distance in terms of redshift for models
alpha, beta, and eta. For comparison, the prediction of the
$\Lambda$CDM model is also shown.} \label{fig:rvsz2ddy}
\end{figure}

The evolution of the quantities $y$, $y'$ and $y''$ in terms of
$z$ are depicted in Figs. \ref{fig:rvsz2y} through
\ref{fig:rvsz2ddy}, for the three models. We have used different
values of the parameters $w$, $\alpha$, $\beta$, and $\eta$, and
assumed the standard values $\Omega_{m0}=0.3$ and $\Omega_{\phi
0}=0.7$. Also, the said evolutions of $y$, $y'$ and $y''$ are
compared with the flat $\Lambda$CDM model with the quoted values
of $\Omega_{m0}$ and $\Omega_{\phi 0}$.

\begin{figure}[th]
\includegraphics[width=6.in,angle=0,clip=true]{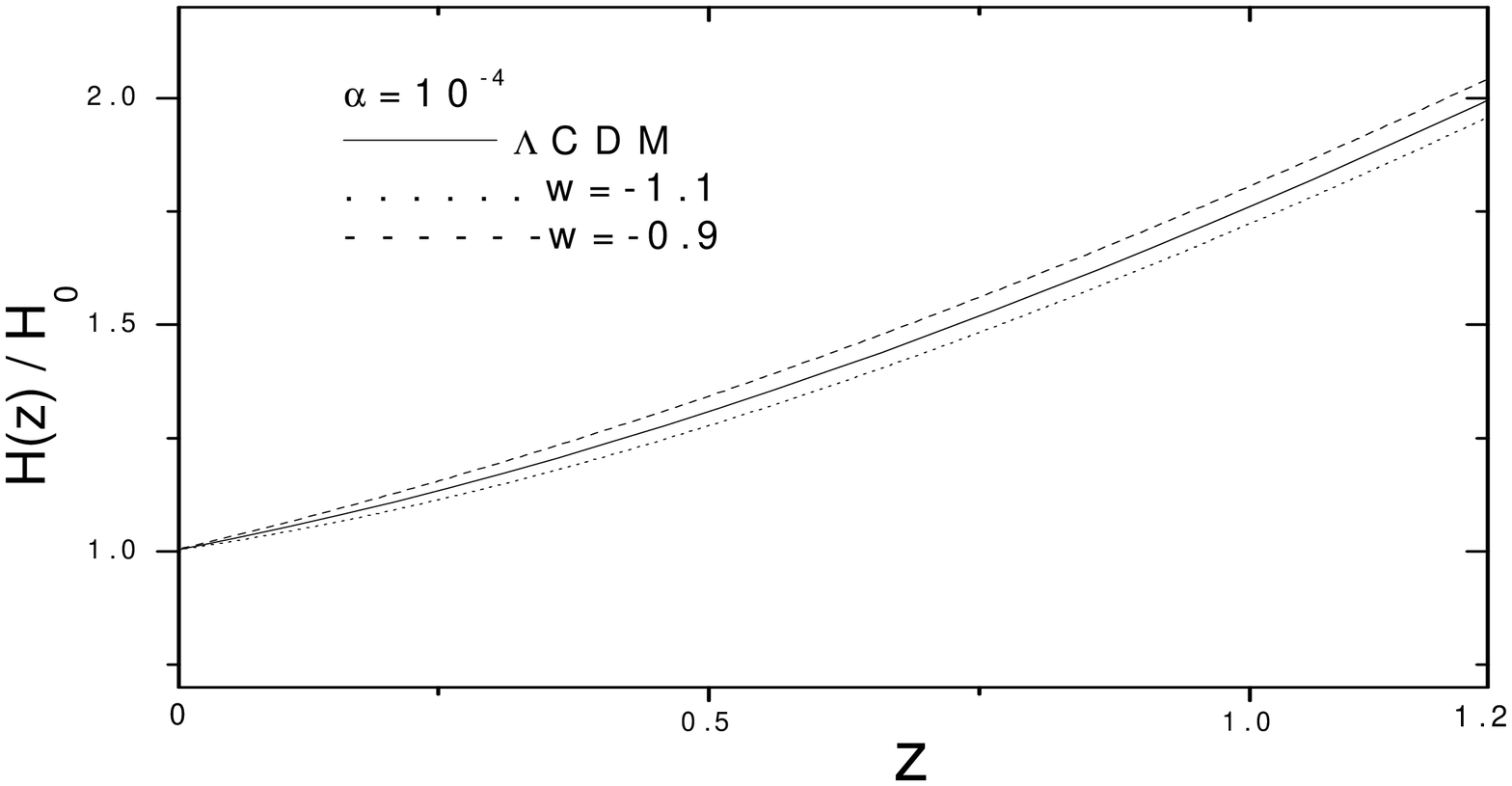}\vspace{-5.cm}
{\vspace{-3.cm}\includegraphics[width=6.in,angle=0,clip=true]{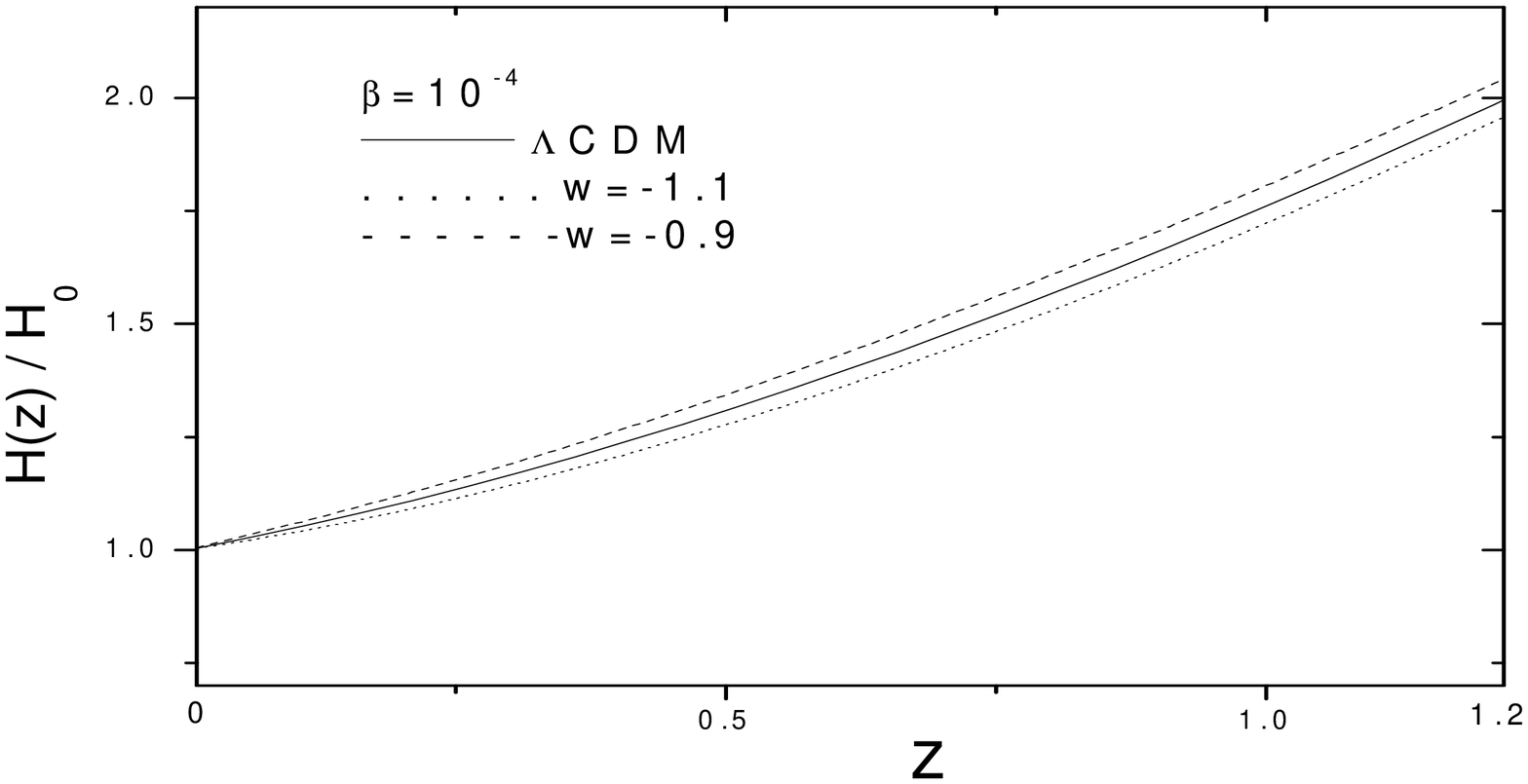}}\vspace{-2.cm}
{\vspace{-4.cm}\includegraphics[width=6.in,angle=0,clip=true]{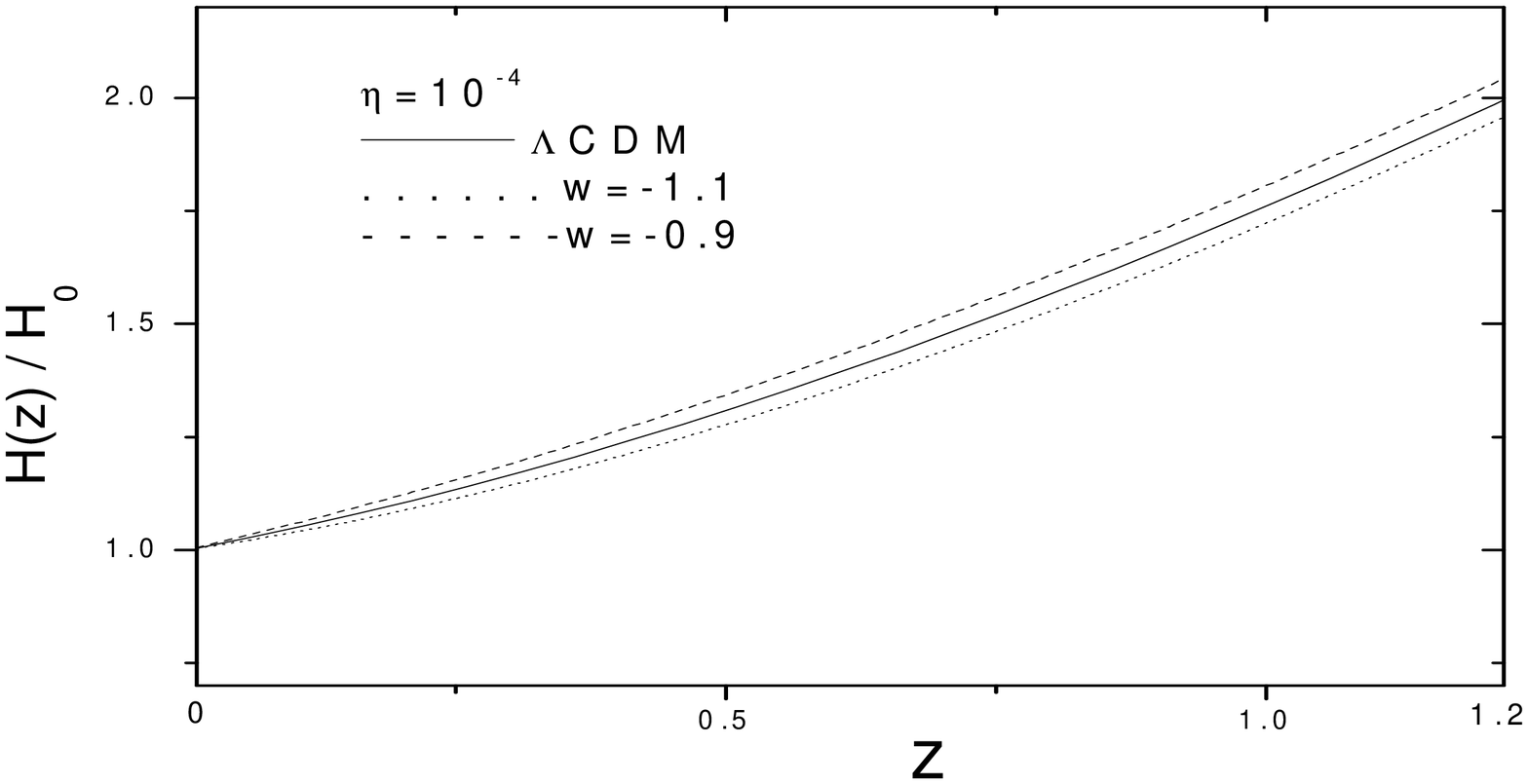}}
\caption{From top to bottom, evolution of the Hubble factor for
models alpha, beta, and eta. For comparison, the prediction of the
$\Lambda$CDM model is also shown.} \label{fig:rvsz2H}
\end{figure}

\begin{figure}[th]
\includegraphics[width=6.in,angle=0,clip=true]{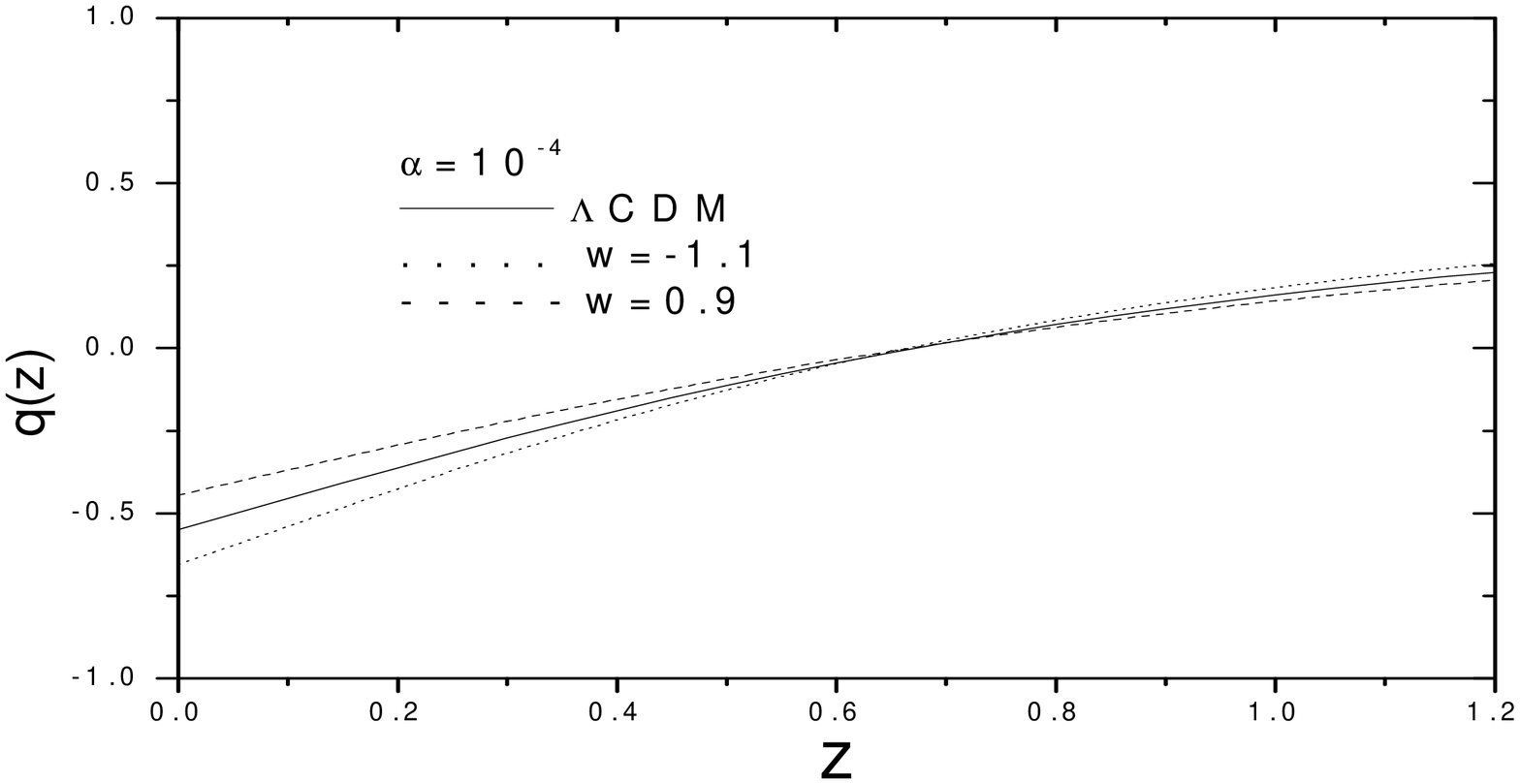}\vspace{-4.3cm}
{\vspace{-3.cm}\includegraphics[width=6.in,angle=0,clip=true]{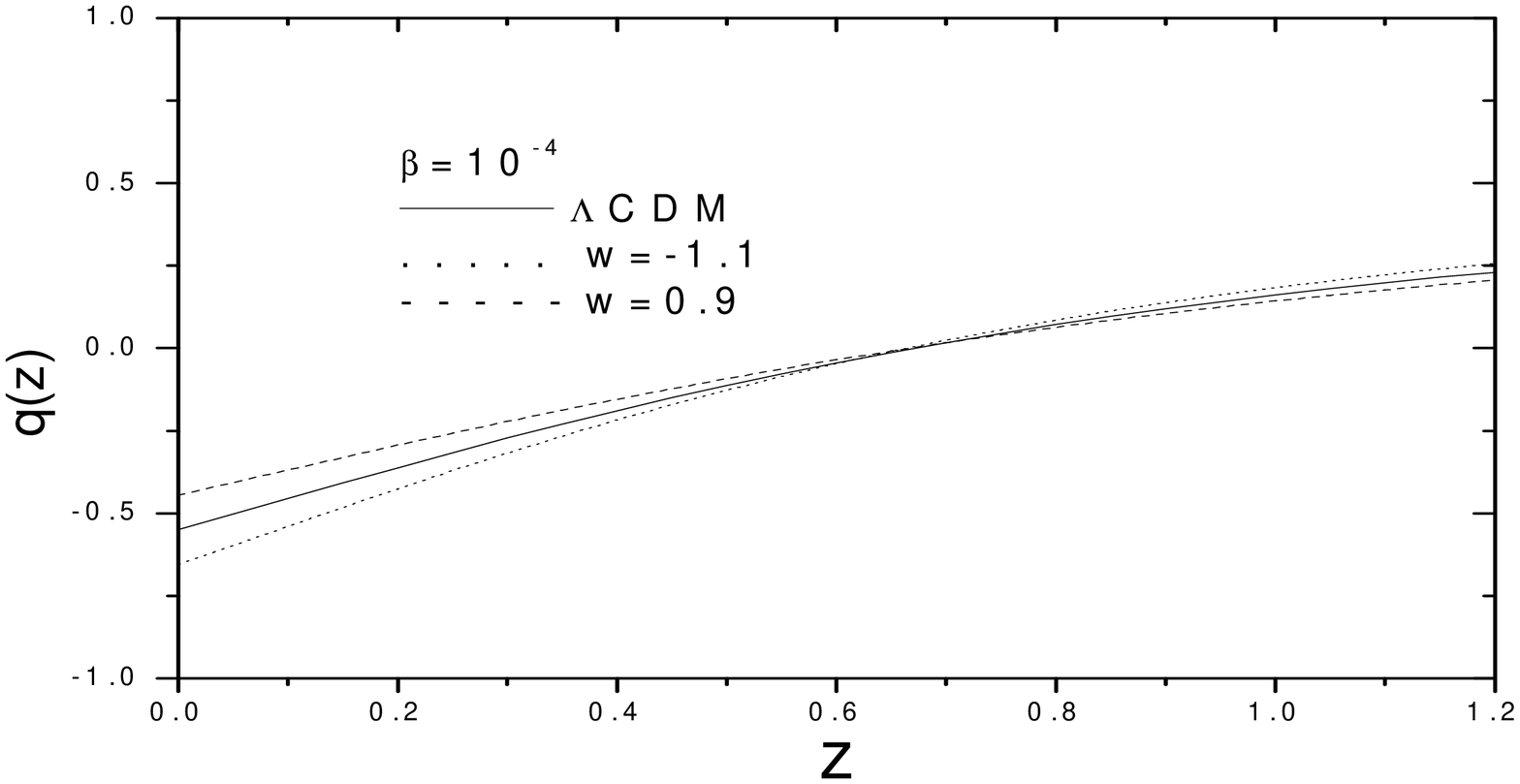}}\vspace{-1.4cm}
{\vspace{-4.cm}\includegraphics[width=6.in,angle=0,clip=true]{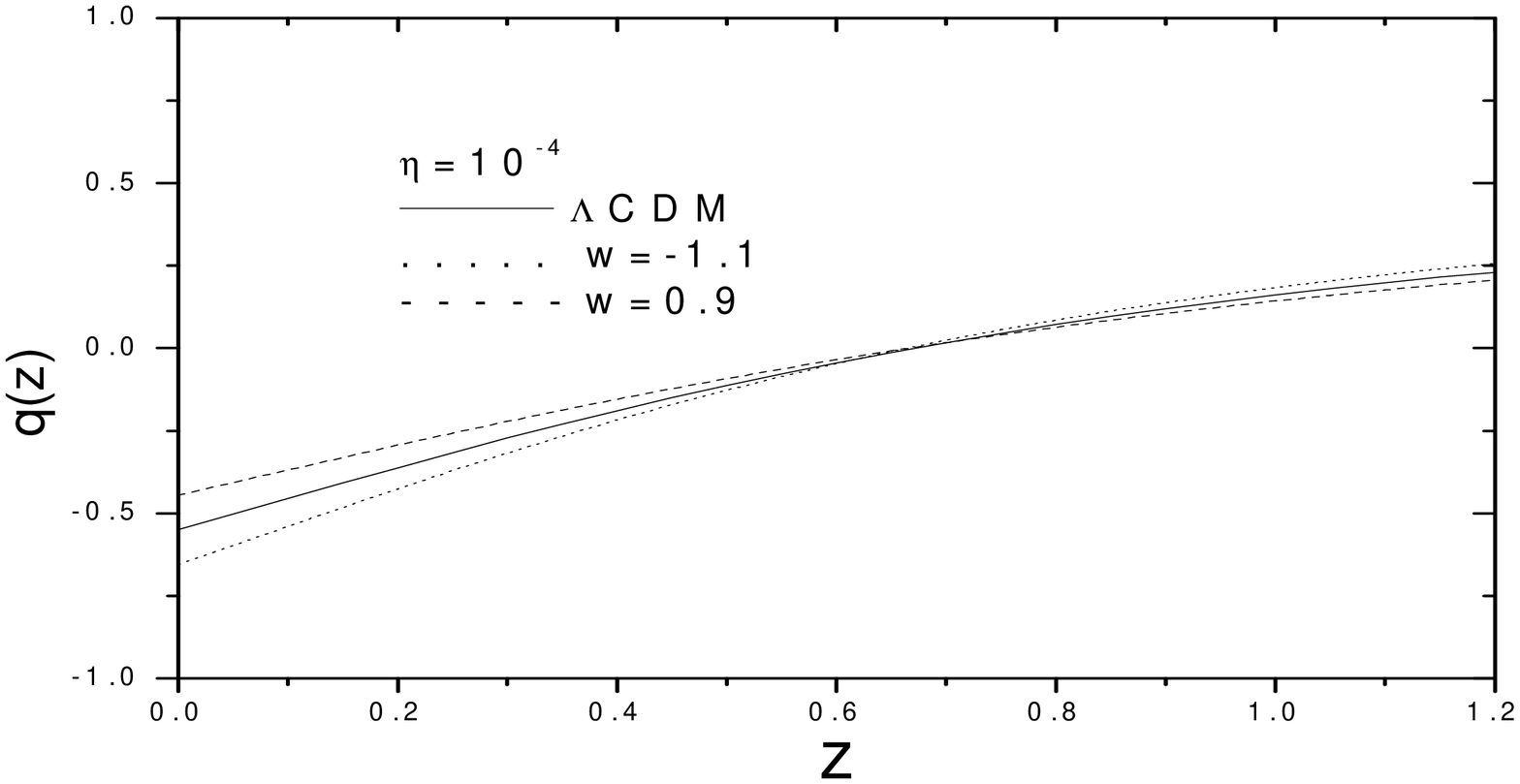}}
\caption{From top to bottom, evolution of the deceleration
parameter  for models alpha,  beta, and eta. For comparison, the
prediction of the $\Lambda$CDM model is also shown.}
\label{fig:rvsz2q}
\end{figure}

\begin{figure}[th]
\includegraphics[width=7.in,angle=0,clip=true]{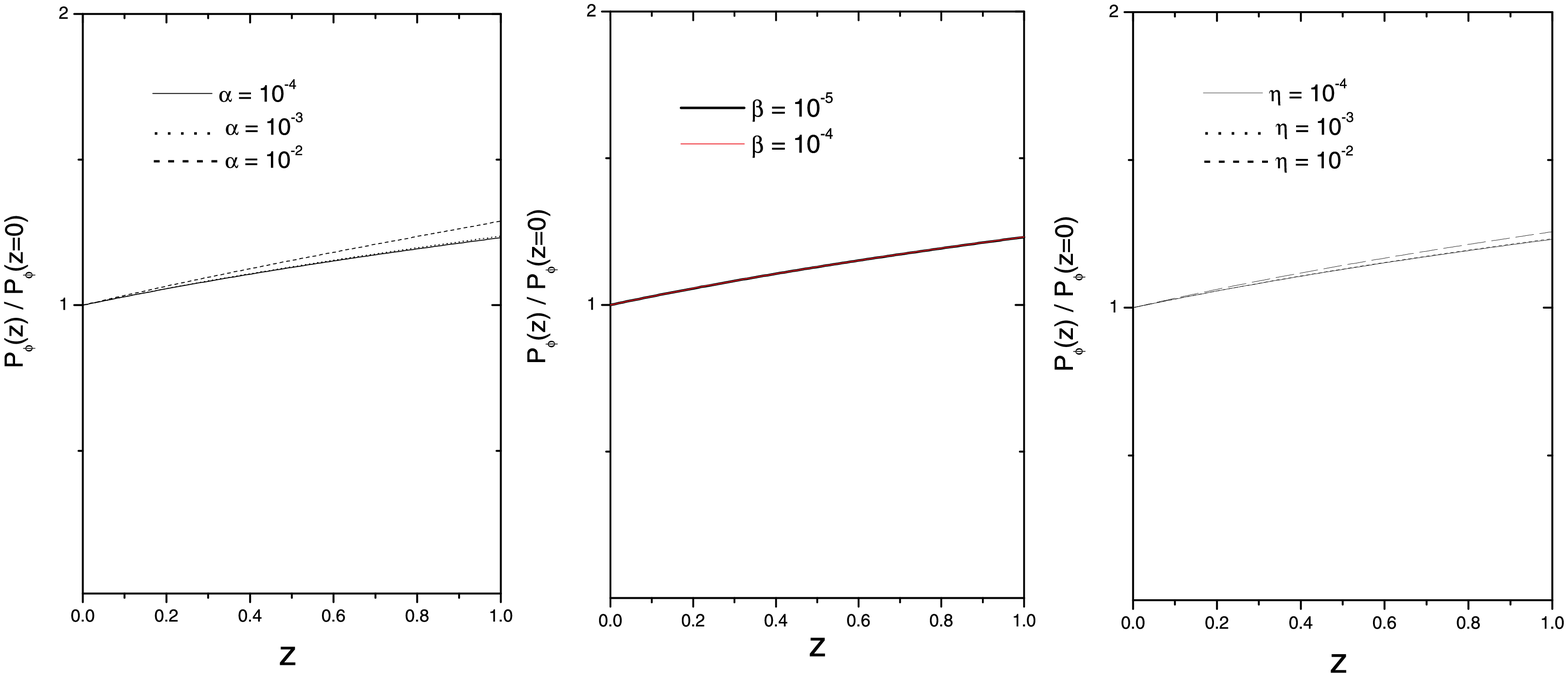}
\caption{From left to right, evolution of the ratio
$P_\phi(z)/P_\phi(z = 0)$ with redshift for models alpha, beta,
and eta.} \label{fig:rvsz2P}
\end{figure}

The data and analysis from the values of $y(z)$, $y'(z)$ and
$y''(z)$ are used to obtain $E(z)$, $q(z)$ and
$P_\phi(z)/P_\phi(z=0)$. The results are shown in Figs.
\ref{fig:rvsz2H} through \ref{fig:rvsz2P} and, again, contrasted
with the standard $\Lambda$CDM model.

Comparison of Figs. \ref{fig:rvsz2y} through \ref{fig:rvsz2q} with
the corresponding figures of Ref. \cite{daly1} reveals that the
three models, alpha, beta, and eta, are consistent with the
analysis of Daly {\it et al.} which is based on different samples
of SN Ia and 30 RG. It is worthy of mention that, except for
adopting the Friedmann-Robertson-Walker metric, the study is
independent of any cosmological assumption. We, therefore,
conclude this section by stressing that the three models are
consistent with the known properties of dark energy.

\section{A strategy to solve the coincidence problem}
As mentioned above, a cosmological model that predicts a constant
or slowly evolving ratio $r$ at late times certainly alleviates
the coincidence problem but this is not enough. The model also
must predict that $r_{0} \sim {\cal O}(1)$. Our strategy is very
simple; it rests on the natural assumption that $r(z)$ and $H(z)$
must be closely interrelated. Therefore, if one express $r$ in
terms of $H$ the problem of explaining why  $r_{0}$ is of order
unity reduces to the less pressing problem of explaining the
present value of $H$. This is interesting because while we are not
aware of model-independent data of $r$ at different redshifts, two
sets of observational data values -though scarce- of $H$ at
various redshifts are now available. One data set was compiled by
computing differential ages of passively evolving galaxies in the
redshift range $0.1 < z < 1.8 \;$ \cite{simon}. The other data set
is based on measurements of supernovae type Ia (SN Ia) and 30
radio galaxies up to redshift $1.2$ \cite{daly1}.

To express $r$ as a function of $H$ we use the identity $\dot{r} =
\dot{H}\, dr/dH$, with
\begin{equation}
\dot{H} = -\frac{\kappa^{2}}{2}\, (\rho_{m} + \rho_{\phi} +
P_{\phi}) = -\frac{3}{2}\; \frac{1+w+r}{1+r}\; H^{2} \, ,
\label{dotH}
\end{equation}
\\
in Eq. (\ref{rr}). Then, the latter becomes
\\
\begin{equation}
\frac{dr}{dH} = \frac{\Im}{H} \, ,
\label{rrH}
\end{equation}
where
\begin{equation}
\Im=-2\,r\,\frac{1+r}{1+w+r}\left[w+\frac{\kappa^{2}\,Q}{9\,H^3}\frac{(r+1)^2}{r}
\right]\, . \label{F}
\end{equation}

Because of  we expect $dr/dH>0$ (a very reasonable assumption
since $H(t<t_{0}) > H_{0}$ as observation tells us) the inequality
\begin{equation}
|w|>\frac{\kappa^{2}\,Q}{9\,H^3}\frac{(r+1)^2}{r} \, \label{cond}
\end{equation}
is to be satisfied.

We note in passing that for the particular case of the
$\Lambda$CDM model (i.e., $w = -1 \, $ and $Q = 0$) one has $\, H
= H_{0} \,[(1+r)/(1+r_{0})]^{1/2}$.

\subsection{Model alpha}
For this model, Eq. (\ref{F}) reduces to
\begin{equation}
\Im = -2 \,r\; \frac{1+r}{1+w+r}\; \left[w+\alpha\frac{(r+1)^2}{r}
\right] \, , \label{F1}
\end{equation}
where we used Eq. (\ref{Fried1}). The condition given by Eq.
(\ref{cond}) leads to restriction $|w|>\alpha\,(1+r)^2/r$.

Using last equation in (\ref{rrH}) and integrating, the Hubble
function can be expressed in terms of $r$ as
\\
\begin{equation}
H(r) = H_{0}\,\exp{[I(r)-I(r_{0})]} \label{HH},
\end{equation}
\\
where $I(r)$ stands for the real part of $\tilde{I}(r)$ with
\\
\begin{equation}
\tilde{I}(r)=\frac{1}{2}\ln(1+r)-\frac{1}{4}\ln[-w\, r
-\alpha(1+r)^2]-\, \frac{1}{2}\frac{w+2}{\sqrt{-w(4\alpha + w)}}
\;\tan^{-1}\left[\frac{2\alpha(1+r)+ w}{\sqrt{-w \,(4\alpha \,
w)}}\right]\, . \label{j2}
\end{equation}

\begin{figure}[th]
\includegraphics[width=3.5in,angle=0,clip=true]{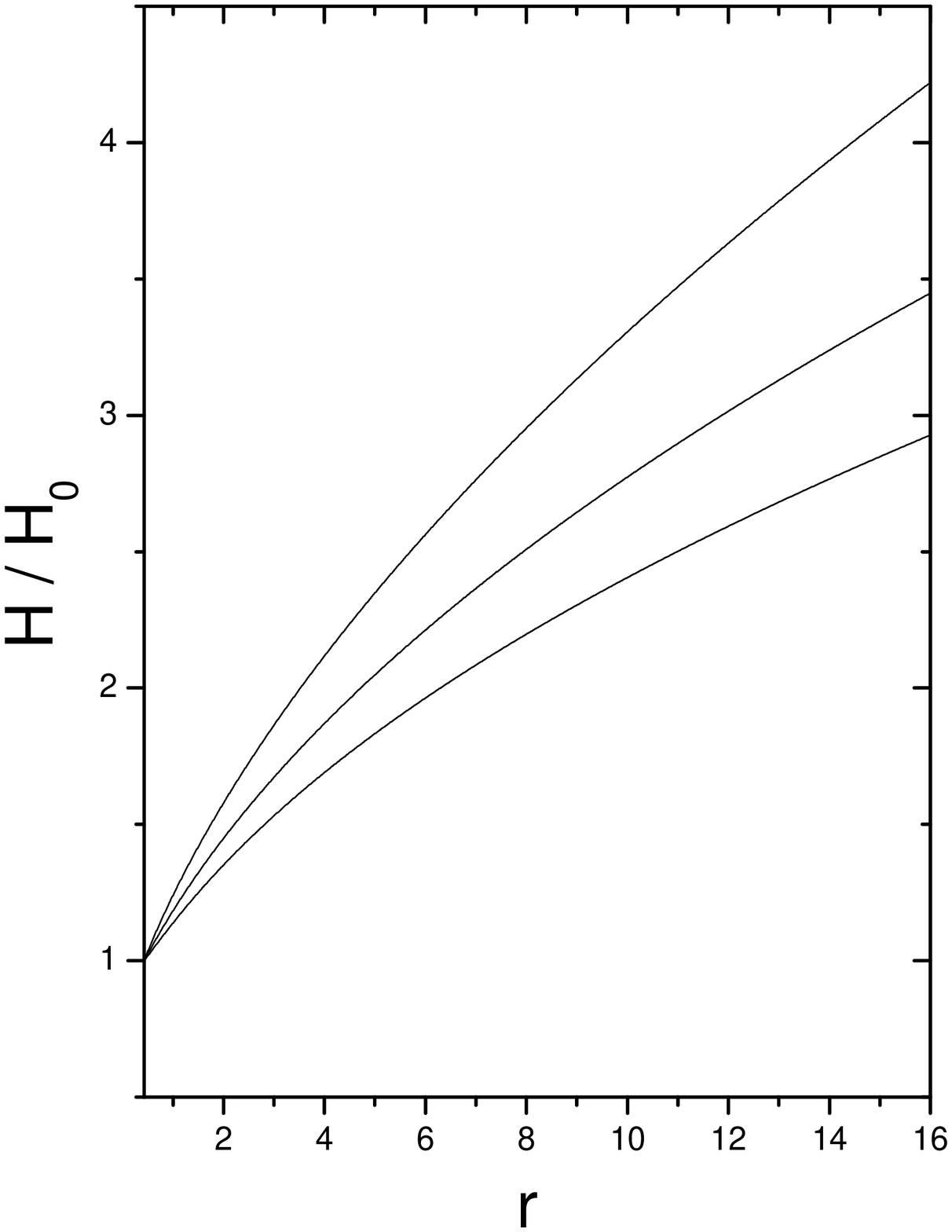}
\caption{Evolution $H$ vs the ratio $r=\rho_{m}/\rho_{\phi}$ as
given by Eq. (\ref{HH}) -model alpha- with $\alpha = 10^{-4}$ for
$w= -0.9$ (quintessence, top line) and $w= -1.1$ (phantom, bottom
line). Also shown is the prediction of the $\Lambda$CDM model
(middle line). In drawing the curves we have set $\, r_{0} =
3/7$.} \label{fig:ratio1l1}
\end{figure}

Figure \ref{fig:ratio1l1} shows the dependence of the Hubble
function on the densities ratio for $\alpha = 10^{-4}$ and two
values of the equation of state parameter, $w$. Likewise, we have
plotted the prediction of the $\Lambda$CDM model. In all the cases
$r_{0} = 3/7$. For other $\alpha$ values compatible with the
restriction $\alpha < 2.3 \times 10^{-3}$ set by the Wilkinson
Microwave Anisotropy Probe 3 yr experiment the corresponding
graphs (not shown) are rather similar.

Unfortunately, as said above, no model-independent  data of $r$ at
different values of $H$ (or $z$) exists whereby, for the time
being, we cannot directly contrast this prediction with
observation. This is why we turn to determine the dependence of
the Hubble factor with redshift and compare it with the two
available observational data sets of $H$ vs $z$. The first
model-independent data set was obtained from the study of the
differential ages of 32 -carefully selected- passively evolving
galaxies in the redshift range $0.1 < z < 1.8 \;$ \cite{simon}.
The age of each galaxy was found by constraining the age of its
older stars with the use of synthetic stellar population models.
The differential ages roughly yields $dz/dt$, then $H(z)$ is given
by   $H = -dz/[(1+z)\,dt]\,$ -see right panel of Fig. 1 in Ref.
\cite{simon}. The second model-independent data set (see lower
panels in Fig. 9 of Daly {\it et al.} \cite{daly1}) was obtained
by applying the  model-independent analysis method of Ref.
\cite{djorgovski} to the coordinate distances of $192$  SN Ia of
Davis {\it et al.} \cite{davis} and $30$ radio galaxies of Ref.
\cite{daly2}.

To obtain the expression for H(z) of this model we combine Eq.
(\ref{HH}) with the integral of (\ref{rr}) in terms of $z$ which
is
\begin{equation}
r(z) =  {\rm Re} \left\{\frac{1}{2\alpha}\left[-w+\gamma \tan
\left( A -\frac{3}{2} \, \gamma \, \ln(1+z)\right) \right]
\right\} -1 \, , \label{rz1}
\end{equation}
where ${\rm Re} \,$  specifies the real part of the quantity in
curly parenthesis, $\, A = \tan^{-1} [(w + 2\alpha
(1+r_{0}))/\gamma]\,$, and $\; \gamma = \sqrt{- w \, (w + 4 \,
\alpha)}$.

\begin{figure}[th]
\includegraphics[width=4.5in,angle=0,clip=true]{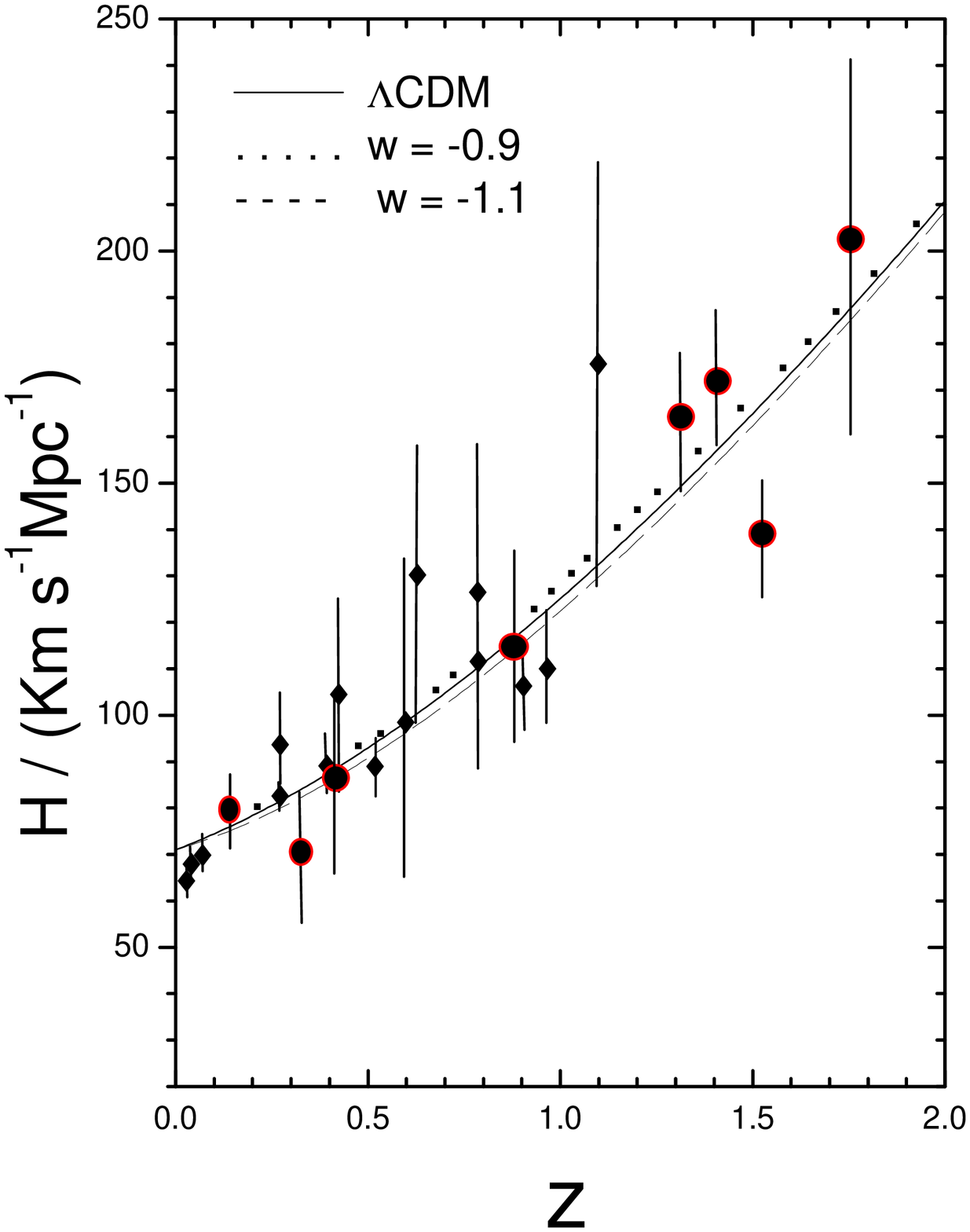}
\caption{Evolution $H$ vs $z$ with $\alpha = 10^{-4}$ for $w=
-0.9$ (quintessence) and $w= -1.1$ (phantom). Also shown is the
prediction of the $\Lambda$CDM model (solid line). In all the
cases we have fixed $r_{0} = 3/7\, $ and $H_{0} = 71 \, {\rm
km/s/Mpc}$. The data points with their $1\sigma$ error bars are
borrowed from Simon {\it et al.} Ref. \cite{simon} (full circles)
and table 2 of Ref. \cite{daly1} (full diamonds).} \label{fig:hz1}
\end{figure}

\begin{figure}[th]
\includegraphics[width=7.5in,angle=0,clip=true]{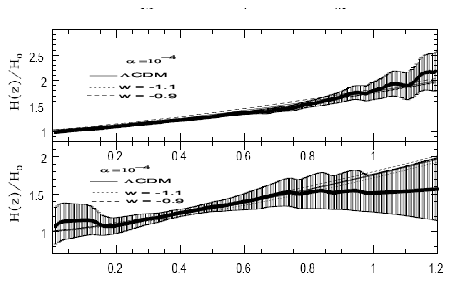}
\vspace*{-3cm}\caption{Same as Fig. \ref{fig:hz1}. The upper panel
shows the combined sample of $132$ SN Ia of Ref. \cite{davis} and
$30$ radio galaxies of Ref. \cite{daly2}. The bottom panel shows
the $30$ radio galaxies only. The data points with their $1\sigma$
error intervals and the best-fit curve (big solid line) in both
panels are borrowed from Daly {\it et al.}, Ref. \cite{daly1}.}
\label{fig:hz2}
\end{figure}

Figures \ref{fig:hz1} and \ref{fig:hz2} show the dependence of the
Hubble expansion rate on redshift predicted by the model for
$\alpha = 10^{-4}$ and two values of the equation of state
parameter, $w$. For comparison, we have also plotted the
prediction of the $\Lambda$CDM model. The data in Fig.
\ref{fig:hz1} are borrowed from Simon {\it et al.} \cite{simon}
(full circles) and Daly {\it et al.} (full diamonds), and in Fig.
\ref{fig:hz2} from Daly {\it et al.} (cf. lower panels of figure 9
in Ref. \cite{daly1}). As can be seen in both figures, the fit to
data does not vary significantly between graphs. Given the
scarcity of data  it is not worthwhile to compare the $\chi^{2}$
of the different curves.

Figures \ref{fig:ratio1l1}, \ref{fig:hz1}, and \ref{fig:hz2} taken
together suggest that this model may be a solution to the
coincidence problem. It fits the available data sets of the
evolution of the Hubble factor not less well than the $\Lambda$CDM
model. Unfortunately, due to the large 1$\sigma$ errors
(especially for $z > 0.5$) they cannot discriminate between these
two models. A like situation occurs for the beta and eta models
detailed below. In consequence, more abundant and accurate
observational information about the aforesaid evolution is needed
in order decide whether any of the three models can really  solve
the coincidence problem.

\subsection{Model beta}
This model has been considered in \cite{guo} and \cite{amendola}.
Equation (\ref{F}), with the help of (\ref{Fried1}), reduces to
\begin{equation}
\Im = -2\,r \;\frac{1+r}{1+w+r}\, \left[w+\beta\;(r+1) \right]\, .
\label{F21}
\end{equation}
Now, the condition set by Eq.(\ref{cond}) boils down to $|w|
>\beta\,(1+r)$.

In virtue of  Eqs. (\ref{rrH}) and (\ref{F21}) the Hubble function
can be cast as
\begin{equation}
H(r) = H_{0}\,\frac{I(r)}{I(r_{0})} \, , \label{HH2}
\end{equation}
where
\begin{equation}
I(r)=\sqrt{(1+r)}\;\;r^{[-w-1]/[2(\beta + w)]}\; [\beta (1+r)+
w]^{(1-\beta)/[2(\beta + w)]} \; .
\label{j2m}
\end{equation}

Numerically, the dependence of the Hubble expansion rate upon $r$
is very close to that of model alpha for $\beta$ values similar to
$\alpha$. This is why we do not show it here.

The dependence of the Hubble expansion rate on redshift can be
obtained by combining Eq. (\ref{HH2}) with (\ref{r}). It is found
that the corresponding graphs for $H(z)$ are rather similar to
those in Figs. \ref{fig:hz1} and \ref{fig:hz2} whereby we do not
depict them here.

\subsection{Model eta}
This model has been studied in \cite{binw}. In this instance,
Eq.(\ref{F}) becomes
\begin{equation}
\Im = -2\,r \;\frac{1+r}{1+w+r}\, \left[w+\eta\;\frac{(r+1)}{r}
\right]\, , \label{F3}
\end{equation}
and the condition set by Eq.(\ref{cond}) simplifies to $|w|\;r
>\eta\,(1+r)$. With the help of  Eqs.(\ref{rrH}) and (\ref{F3}) the Hubble
function, $H(r)$, is given formally by (\ref{HH2}) but this time
\begin{equation}
I(r)=\sqrt{(1+r)}\;\;[\eta+\,r\,(w+\eta)]^{-\frac{(1+w+\eta)}{2(w+\eta)}}
\; . \label{j3m}
\end{equation}
By combining the expression for $H(r)$ with the integral of
(\ref{rr}) in terms of $z$,
\\
\begin{equation}
r(z) =
\frac{(1+z)^{-3(w+\eta)}}{(w+\eta)}\,\left\{r_0(w+\eta)+\eta\,(1-(1+z)^{3(w+\eta)})
\right\} \, , \label{rz3}
\end{equation}
one follows $H(z)$. Again, the graphs of $H(z)$ are very close to
those of Figs. \ref{fig:hz1} and \ref{fig:hz2} and we do not show
them here.

\section{Discussion}
Nowadays the leading candidate for dark energy is the popular
$\Lambda$CDM model. It is simple, it has just one free parameter
-the vacuum energy density- and fits reasonably well most (if not
all) observational data. However, it presents the puzzle (apparent
or not) that we happen to live at a very special epoch: the
transient epoch at which the ratio between the densities of matter
and vacuum energy is of order unity. To solve this  one must look
for alternative models in which the ratio $r$ stays constant or
varies very slowly, around the present time, with respect to the
Universe expansion. Interacting models (with $Q >0$, see Eq.
(\ref{rr})) possess this property. But it does not suffice. The
successful model must also predict $r_{0} \sim {\cal O}(1)$.

In this paper we considered three interacting models that present
the first characteristic and may also have the second one. To
decide on the latter one must contrast the $r(z)$ function
predicted by each model with observation. Regrettably, as far as
we know, model-independent data about $r(z)$ do not exist.
However, albeit with wide uncertainties, there exist
model-independent data of the evolution of the Hubble factor up to
$z = 1.8$ \cite{daly1,simon}. Therefore, we have numerically
determined $H(r)$ and $H(z)$ for each model and compared the
latter  with these observational data. All three models show
compatibility with the data but only in a  degree similar to the
$\Lambda$CDM model. Future, more accurate and extensive
measurements will likely tell us whether the latter model fits the
data better or worse than interacting models (the ones presented
in this paper or maybe others). If, in the end, the $\Lambda$CDM
model makes the best fit, then we may conclude that the
coincidence problem is not a problem any more; it is simply  a
coincidence and therefore not a problem.

\acknowledgements{S.d.C. was supported from Comisi\'{o}n Nacional
de Ciencias y Tecnolog\'{\i}a (Chile) through FONDECYT Grants No.
1040624, No. 1051086, and No. 1070306 and by the PUCV. R.H. was
supported by the ``Programa Bicentenario de Ciencia y
Tecnolog\'{\i}a" through the Grant ``Inserci\'on de Investigadores
Postdoctorales en la Academia" \mbox {N$^0$ PSD/06}. D.P.
acknowledges ``FONDECYT-Concurso incentivo a la cooperaci\'{o}n
internacional" No. 7070003, and is grateful to the ``Instituto de
F\'{\i}sica" for warm hospitality; also D.P. research was
partially supported by the ``Ministerio Espa\~{n}ol de
Educaci\'{o}n y Ciencia" under Grant No. FIS2006-12296-C02-01.}

\end{document}